\documentclass[fleqn,usenatbib]{mnras}

\usepackage{newtxtext,newtxmath}
\usepackage{xcolor, soul}
\definecolor{teagreen}{rgb}{0.82, 0.94, 0.75}
\sethlcolor{teagreen}

\usepackage[T1]{fontenc}

\DeclareRobustCommand{\VAN}[3]{#2}
\let\VANthebibliography\thebibliography
\def\thebibliography{\DeclareRobustCommand{\VAN}[3]{##3}\VANthebibliography}

\usepackage{graphicx}	
\usepackage{amsmath}	

\usepackage{stackengine}

\usepackage{caption}

\title[WISDOM -- XXIII.
SFEs of bulges]{WISDOM project -- XXIII. 
Star-formation efficiencies of eight early-type galaxies and bulges observed with SITELLE and ALMA}

\author[A.\ Lu et al.]{Anan Lu,$^{1}$\thanks{E-mail: anan.lu@mail.mcgill.ca}
Daryl Haggard,$^{1}$
Martin Bureau,$^{2}$
Jindra Gensior,$^{3,4}$
Carmelle Robert,$^{5}$
Thomas G.\ Williams,$^{2}$
\newauthor
Fu-Heng Liang,$^{2,6}$
Woorak Choi,$^{7,8}$
Timothy A.\ Davis,$^{9}$
Ilaria Ruffa$^{10}$,
Sara Babic,$^{1}$
Hope Boyce,$^{1}$
\newauthor
Michele Cappellari,$^{2}$
Benjamin Cheung,$^{1}$
Laurent Drissen,$^{5}$
Jacob S.\ Elford,$^{9}$
Thomas Martin,$^{5}$
\newauthor
Carter Rhea,$^{11}$
Laurie Rousseau-Nepton$^{12}$,
Marc Sarzi,$^{13}$
Hengyue Zhang$^{2}$
\\
$^{1}$Trottier Space Institute and Department of Physics, McGill University, 3600 University Street, Montreal, QC H3A~2T8, Canada\\
$^{2}$Sub-department of Astrophysics, Department of Physics, University of Oxford, Denys Wilkinson Building, Keble Road, Oxford OX1~3RH, UK\\
$^{3}$Institute for Astronomy, University of Edinburgh, Royal Observatory, Blackford Hill, Edinburgh EH9 3HJ, UK\\ 
$^{4}$Department of Astrophysics, University of Zurich, Winterthurerstrasse 190, 8057 Z{\"u}rich, Switzerland\\
$^{5}$D{\'e}partement de Physique, de G{\'e}nie Physique et d’Optique, Universit{\'e} Laval, Qu{\'e}bec, QC G1V~0A6, Canada\\
$^{6}$European Southern Observatory (ESO), Karl-Schwarzschild-Stra{\ss}e 2, 85748 Garching, Germany\\
$^{7}$Department of Physics and Astronomy, McMaster University, Hamilton, ON L8S 4M1, Canada\\
$^{8}$Department of Astronomy, Yonsei University, 50 Yonsei-ro, Seodaemun-gu, Seoul 03722, Republic of Korea\\
$^{9}$Cardiff Hub for Astrophysics Research \& Technology, School of Physics \& Astronomy, Cardiff University, Queens Buildings, The Parade, Cardiff, CF24~3AA,UK\\
$^{10}$INAF, Arcetri Astrophysical Observatory, Largo Enrico Fermi 5, I-50125 Florence, Italy\\
$^{11}$D{\'e}partement de Physique, Universit{\'e} de Montr{\'e}al, Succ.\ Centre-Ville, Montr{\'e}al, Qu{\'e}bec, H3C~3J7, Canada \\
$^{12}$David A.\ Dunlap Department of Astronomy and Astrophysics, University of Toronto, 50 St-George Street, Toronto, M5S~3H4, Canada
\\
$^{13}$Armagh Observatory and Planetarium, College Hill, Armagh BT61 9DG, UK\\
}

\date{Accepted XXX. Received YYY; in original form ZZZ}

\pubyear{XXX}

\begin{document}
\label{firstpage}
\maketitle

\begin{abstract} 
  Early-type galaxies (ETGs) are known to harbour dense spheroids of stars with scarce star formation (SF). Approximately a quarter of these galaxies have rich molecular gas reservoirs yet do not form stars efficiently. These gas-rich ETGs have properties similar to those of bulges at the centres of spiral galaxies. We use spatially-resolved observations ($\sim100$~pc resolution) of warm ionised-gas emission lines (H$\beta$, [\ion{O}{iii}], [\ion{N}{ii}], H$\alpha$ and [\ion{S}{ii}]) from the imaging Fourier transform spectrograph SITELLE at the Canada-France-Hawaii Telescope and cold molecular gas (\textsuperscript{12}CO(2-1) or \textsuperscript{12}CO(3-2)) from the Atacama Large Millimeter/submillimeter Array (ALMA) to study the SF properties of $8$ ETGs and bulges. We use the ionised-gas emission lines to classify the ionisation mechanisms and demonstrate a complete absence of regions dominated by SF ionisation in these ETGs and bulges, despite abundant cold molecular gas. The ionisation classifications also show that our ETGs and bulges are dominated by old stellar populations. We use the molecular gas surface densities and H$\alpha$-derived SF rates (in spiral galaxies outside of the bulges) or upper limits (in ETGs and bulges) to constrain the depletion times (inverse of the SF efficiencies), suggesting again suppressed SF in our ETGs and bulges. Finally, we use the molecular gas velocity fields to measure the gas kinematics, and show that bulge dynamics, particularly the strong shear due to the deep and steep gravitational potential wells, is an important SF-regulation mechanism for at least half of our sample galaxies. 

\end{abstract}


\section{Introduction}
\label{intro}

Local galaxies (redshifts $z\lesssim0.1$) can be roughly divided into two classes \citep[e.g.][]{2000AJ....120.1579York}: blue and thus star-forming mostly late-type galaxies (LTGs), forming a so-called star-forming `main sequence', and red and thus quiescent mostly early-type galaxies (ETGs), with lower star-formation rates (SFRs; e.g.\ \citealt{kauffmann_2003MNRAS.341...54K, Cirasuolo2007MNRAS.380..585C}).  
Most galaxies harbour a spheroid of old stars (i.e.\ a bulge) at their centre, which is more prominent in ETGs. As shown in e.g.\ \citet{2003MNRAS.345.1381Ferreras}, \citet{2005MNRAS.358.1477Athanassoula} and \citet{2022MNRAS.513..256Dimauro}, bulges have typically experienced mergers and intense star formation (SF) in the early universe, but have low current SFRs. 
This leads to current specific SFRs (sSFRs; i.e.\ SFRs per unit stellar mass, ${\rm SFR}/M_\star$) that are much smaller than those of LTGs. 

The cause of this SF quenching in ETGs and bulges remains uncertain. One historical solution is to associate low SF with a lack of cold (molecular) gas, that can be depleted in ETGs and bulges via several mechanisms such as merger-triggered starbursts and active galactic nucleus (AGN) feedback (see e.g.\ \citealt{2018NatAs...2..695Man} and references therein). 
However, recent molecular gas surveys of nearby galaxies show that $\approx23\%$ of ETGs in the local universe have a substantial molecular gas reservoir \citep{Young2011MNRAS.414..940Y, Davis2019MNRAS.486.1404D, Ruffa2019MNRAS.484.4239R}, indicating that SF in many ETGs must instead be quenched by low SF efficiencies (SFEs; i.e.\ SFRs per unit molecular gas mass, ${\rm SFR}/M_{\rm mol}$). It is thus useful to quantify the timescale required to exhaust the molecular gas at the current SFR, known as the molecular gas depletion time ($\tau_{\rm dep}\equiv{\rm SFE}^{-1}$). The $\tau_{\rm dep}$ of starburst galaxies is $10$ -- $100$~Myr, but it is much longer for spiral galaxies, reaching $1$ -- $2$~Gyr \citep{kennicutt2012star,Leroy2008AJ....136.2782L}. This is much longer than the typical free-fall time of the giant molecular gas clouds \citep{2007ARA&A..45..565McKee} where most of the molecular gas resides. 
The $\tau_{\rm dep}$ of gas-rich ETGs and bulges are even longer than those of LTGs \citep{2013ApJ...778....2Saintonge,Davis2014MNRAS.444.3427D}, inviting further investigation of SF quenching mechanisms. SF in the bulges of LTGs, such as in the Central Molecular Zone of our own Milky Way \citep[e.g.][]{Davis2014MNRAS.444.3427D,2014MNRAS.440.3370Kruijssen}, is also quenched, especially that in the dense gas. Thus, it is natural to combine samples of bulges and gas-rich ETGs to search for an explanation of the quenched SF in those environments.

One useful tool to probe SFEs is the scaling relation between the cold gas mass surface density ($\Sigma_{\rm gas}$) and the SFR surface density ($\Sigma_{\rm SFR}$), originally proposed by \citet{kennicutt1998global} on galactic scales. Using a sample of star-forming galaxies observed on kpc scales, \citet{bigiel2008star} established that $\Sigma_{\rm SFR}$ is more tightly correlated with the molecular gas mass surface density ($\Sigma_{\rm mol}$) than with the total cold gas mass surface density, indicating that the molecular gas depletion time is key to the study of SF regulation mechanisms. The ALMA-MaNGA QUEnching and STar formation survey (ALMaQUEST; \citealt{2020ApJ...903..145Lin, 2021MNRAS.501.4777Ellison}) examined over $50$ different types of galaxies on kpc scales and revealed significant variations of $\tau_{\rm dep}$ across different environments (e.g.\ galaxy type and/or region within a galaxy). Approaching the $100$~pc scale, \citet{2021A&A...650A.134Pessa} revealed that the usual kpc-scale power-law relation between $\Sigma_{\rm SFR}$ and $\Sigma_{\rm mol}$ still holds true, despite a large scatter. This in turn suggests that the global scaling relation is an outcome of the local mechanisms that drive SF, while the scatter offers insight into the mechanisms dominating SF in different environments. There is, however, a lack of $100$~pc-scale $\tau_{\rm dep}$ studies of ETGs.

Spatially resolving the SFEs of individual galaxies also allows to probe e.g.\ the radial dependences of the depletion times. Based on kpc-scale studies, \citet{2021ApJ...923...60Vcallifa} showed that the SFEs are enhanced toward the centres of nearby spiral galaxies. 
However, for galaxies with SFRs lower than those of main-sequence galaxies, the SFEs are suppressed in the centres \citep{2024ApJ...964..120PanAQ}. The decrease of the SFEs toward the centres of these galaxies is associated with a dramatic decrease of the SFRs but constant gas fractions, thus suggesting that the low(er) SFRs of these galaxies are related to less efficient conversion of gas into stars, rather than gas depletion or expulsion. 
At $100$~pc scale, it is more difficult to extract radial profiles of SFEs, as \ion{H}{ii} regions and molecular clouds are often not co-spatial. Nevertheless, ETGs typically have smooth distributions of molecular gas and SFR, enabling robust analyses of their azimuthally-averaged SFE radial profiles at $100$~pc scale \citep{2022MNRAS.514.5035Lu,2024MNRAS.531.3888Lu}. Furthermore, $100$~pc-scale analyses of the SFEs, especially of the radial profiles, offer insight into the transition from star-forming (or even star-bursting) regions in the galaxy discs to the quenched SF in the bulges. In addition, that same spatial decorrelation of \ion{H}{ii} regions and molecular clouds can be used to characterise the relevant gas timescales \citep{2019Natur.569..519Kruijssen}.

In this paper, we probe the depletion times (inverse of the SFEs) of $8$ ETGs and spiral galaxy bulges at $100$~pc scale, combining high-spatial resolution CO (a cold molecular gas tracer) measurements from the Atacama Large Millimeter/submillimeter Array (ALMA) and ionised hydrogen (a SFR indicator) measurements from the imaging Fourier transform spectrograph SITELLE at the Canada-France-Hawaii Telescope (CFHT). In Section~\ref{data}, we detail the processes used to collect and calibrate the SITELLE and ALMA data, convert to surface density maps and calculate depletion times. In Section~\ref{results}, we classify the ionisation mechanisms at $100$~pc scale and probe the $\Sigma_{\rm SFR}$ -- $\Sigma_{\rm mol}$ relations and $\tau_{\rm dep}$ radial profiles. In Section~\ref{discussion}, we discuss the measured SFEs and relate them to potential SF regulation mechanisms. We also discuss the uncertainties and potential biases. We conclude in Section~\ref{conclusion}.

\section{Data and Methods}
\label{data}

\subsection{Targets}
\label{data:targets}

In this work we investigate $8$  galaxies observed as part of the mm-Wave Interferometric Survey of Dark Object Masses (WISDOM) project, for which we have obtained both ALMA and SITELLE data to probe the molecular and the ionised gas, respectively. The ALMA data collected by WISDOM were originally obtained with the intent to measure supermassive black hole (SMBH) masses through kinematic modelling of the molecular gas. The sample is thus fairly heterogeneous, containing both nearly-quenched ETGs and star-forming spirals, and galaxies with active and inactive nuclei. The sub-sample selected to be observed with SITELLE is composed of 5 ETGs and 3 spiral galaxies with large bulges. They all have high molecular gas mass surface densities that should enable SF, but appear to have low SFRs based on integrated {\it Galaxy Evolution Explorer} ({\it GALEX}) far-ultraviolet (FUV) and {\it Wide-field Infrared Survey Explorer} ({\it WISE}) $22$-$\mu$m observations. These galaxies are mostly nearby ($\lesssim20$~Mpc distance), so that it is possible to probe their SFEs at $100$~pc scale using $\approx1\arcsec$ resolution observations with SITELLE. However, two of our ETG targets are more distant (NGC~383 and NGC~708, at distances of $\approx70$~Mpc), so that we can only measure the SFEs at a physical resolution of $\approx300$~pc in those two objects. Nevertheless, we can still spatially resolve the molecular gas and ionised-gas discs.  
The large spirals (NGC~3169 and NGC~4501) and an interacting galaxy pair (NGC~4438 and NGC~4435) in our sample can be uniquely captured by SITELLE's large field of view (FOV) without mosaicing. We summarise the properties of the sample galaxies in Table~\ref{tab1}.

\begin{table*}
\caption{Properties of the sample galaxies.}
\label{tab1}
    \centering
    \resizebox{\textwidth}{!}{%
    \begin{tabular}{|l|c|c|c|c|c|c|c|c|c|l}
    \hline 
    Name & Distance & $V_{\rm sys}$ & $1\arcsec$ scale & Type & $\log\left(\frac{M_{\rm mol,tot}}{{\rm M}_\odot}\right)$& $\log\left(\frac{{\rm SFR_{ul,H\alpha}}}{{\rm M}_\odot\,{\rm yr}^{-1}}\right)$ & $\log\left(\frac{{\rm SFR_{ul,FUV+22\mu m}}}{{\rm M}_\odot\,{\rm yr}^{-1}}\right)$ & $\log\left(\frac{M_\star}{{\rm M}_\odot}\right)$ & $\log\left(\frac{\mu_\star}{{\rm M}_\odot\,{\rm kpc}^{-2}}\right)$ &$A_V$\\  
    & (Mpc) & (km~s$^{-1}$) & (pc) & & & & & &  &\\  
    (1)& (2)& (3)& (4)& (5)& (6) &(7)& (8)& (9)& (10) &(11)\\
    \hline 
        NGC~\phantom{0}383 & 66.6 & 5000 & 329 & E & 9.15$~\pm~$0.06& -0.27$~\pm~$0.10 & \phantom{-}0.00& 11.82 & 9.92 &1.6$~\pm~$0.23\\
        NGC~\phantom{0}524 & 23.3 & 2450 & 115& E & 7.99$~\pm~$0.06& -0.40$~\pm~$0.25& -0.56 & 11.40& 9.75 &2.4$~\pm~$0.62\\
        NGC~\phantom{0}708 & 58.3 & 4900 & 267 & E & 8.55$~\pm~$0.08& -0.28$~\pm~$0.19 & -0.29 & 11.75 & 9.30 &1.5$~\pm~$0.54\\
        NGC~3169 & 18.7 & 1250 & \phantom{1}90 & S & 9.54$~\pm~$0.04& \phantom{-}0.50$~\pm~$0.23& \phantom{-}0.29 & 10.84 & 8.26 &1.2$~\pm~$0.59\\
        NGC~4429 & 16.5 & 1100 & \phantom{1}80 & E & 8.05$~\pm~$0.05& -0.92$~\pm~$0.24 & -0.86 & 11.17 & 9.19 &1.2$~\pm~$0.60\\
        NGC~4435$^\dagger$ & 16.5 & \phantom{1}850 & \phantom{1}80 & E & 8.56$~\pm~$0.05& -0.87$~\pm~$0.28 & -0.84 & 10.69 & 9.18 &1.3$~\pm~$0.66\\
        NGC~4438$^\dagger$ & 16.5 & \phantom{11}70 & \phantom{1}80 & S & 9.45$~\pm~$0.04& -0.37$~\pm~$0.15 & -0.30& 10.75 & 9.42 &2.1$~\pm~$0.26\\
        NGC~4501 & 15.3 & 2280 & \phantom{1}67 & S & 8.71$~\pm~$0.04& \phantom{-}0.47$~\pm~$0.14 & \phantom{-}0.43 & 11.00& 8.94 &1.8$~\pm~$0.25\\
    \hline
    \end{tabular}
    }
{\it Notes:} {(1) Galaxy name. The interacting galaxy pair, NGC~4438 and NGC~4435, is marked with daggers. (2) Distance (NASA Extragalactic Database redshift-independent distance catalogue; \citealt{steer2016redshift}). (3) Systemic velocity, measured from our ALMA CO data. (4) Conversion from angular to physical scale, based on the galaxy distance. (5) Galaxy type: E (ETG), S (spiral). (6) Molecular gas mass with $1\sigma$ uncertainties (from our CO flux measurement), measured within the ALMA FOV (see Section~\ref{data:CO} for details). 
(7) Total SFR with $1\sigma$ uncertainties (from our H$\alpha$ and H$\beta$ flux measurements), measured over the entire galaxy using the SITELLE data. This quantity is formally an upper limit, as it does not separate different ionisation mechanisms (see Section~\ref{data:Halpha} for details). (8) Total SFR, measured over the entire galaxy using the FUV and $22$-$\mu$m data of \citet{Davis2022MNRAS.512.1522D}. The contribution from old stellar population is subtracted from these measurements.  
(9) Total stellar mass \citep{Davis2022MNRAS.512.1522D}. (10) Stellar mass surface density within the effective radius \citep{Davis2022MNRAS.512.1522D}. (11) $V$-band extinction of the galaxy, calculated using spaxels for which both the H$\alpha$ and the H$\beta$ emission have $S/N>3$.}
\end{table*}

\subsection{Molecular gas observations}
\label{data:CO}

In this work, the molecular gas was observed using ALMA as part of the WISDOM project. The observations of the \textsuperscript{12}CO(2-1) and \textsuperscript{12}CO(3-2) emission lines were obtained between 2013 and 2020 as part of a number of programmes: 2013.1.00493.S (PI:Bureau), 2015.1.00466.S (PI: Onishi), 2015.1.00419.S (PI:Davis), 2015.1.00598.S (PI:Bureau), 2016.1.00437 (PI:Davis), 2016.2.00053.S (PI: Liu), 2017.1.00391.S (PI: North) and 2019.1.00582.S (PI:Bureau). Each target was observed multiple times using multiple array configurations with different minimum and maximum baseline lengths, to reach high angular resolutions while ensuring excellent flux recovery. 
The data were calibrated and combined using the {\tt Common Astronomy Software Applications} ({\tt CASA}; \citealt{McMullin2007ASPC..376..127M}) pipeline. To remove the continuum emission from the line spectral window of each dataset, a linear fit was made to the line-free channels at both ends of that window as well as the three other (pure continuum) spectral windows, and the fit was then subtracted in the $uv$ plane using the {\tt CASA} task {\tt uvcontsub}. The resulting line data of each galaxy were then imaged into a RA-Dec.-velocity cube with a (binned) channel width of $10$~km~s$^{-1}$ and a pixel size properly sampling the synthesised beam. 
Briggs' weighting with a robust parameter of $0.5$ and $uv$ tapering were used to achieve a synthesised beam as close as possible to the seeing of the H$\alpha$ observations ($\approx1\farcs0$, see Section~\ref{data:Halpha}). 
The data cubes were cleaned in regions of line emission to a threshold of $\approx1.2$ times the root-mean-square (RMS) noise ($\sigma_{\rm RMS}$) measured from line-free channels. 
We then convolved each data cube spatially with a narrow and slightly elongated two-dimensional Gaussian, to achieve a perfectly circular synthesised beam that exactly matches the seeing of the H$\alpha$ observations. The pixel size was also adjusted to match that of the SITELLE observations ($0\farcs3125\times0\farcs3125$). The zeroth (total intensity) and first (intensity-weighted mean line-of-sight velocity) moment maps of the data cubes were then created using a masked-moment technique and are shown in Figure~\ref{fig:maps}.

\begin{figure*}

\centering\includegraphics[width=0.99\textwidth]{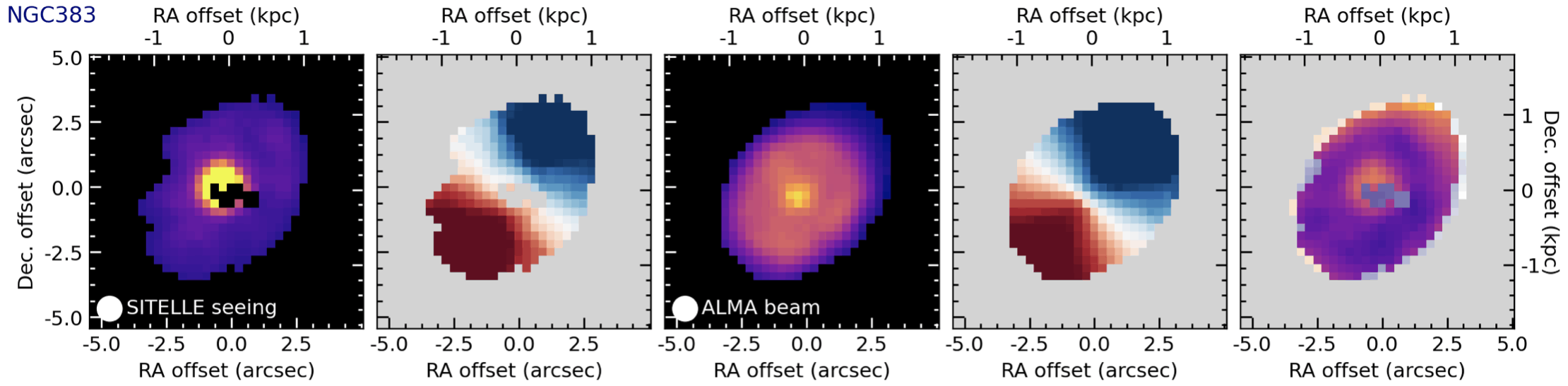}\\\includegraphics[width=0.99\textwidth]{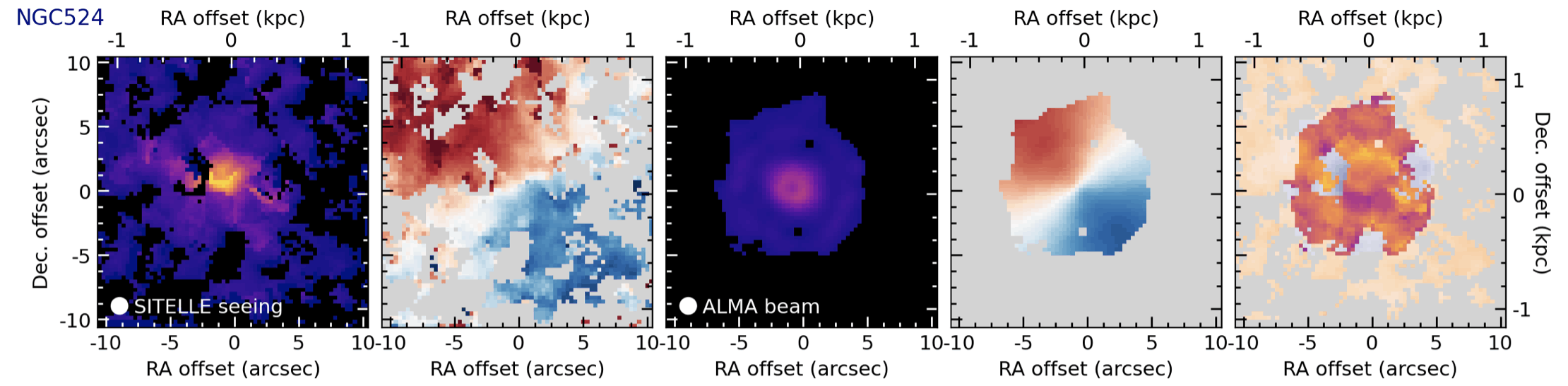}\\\includegraphics[width=0.99\textwidth]{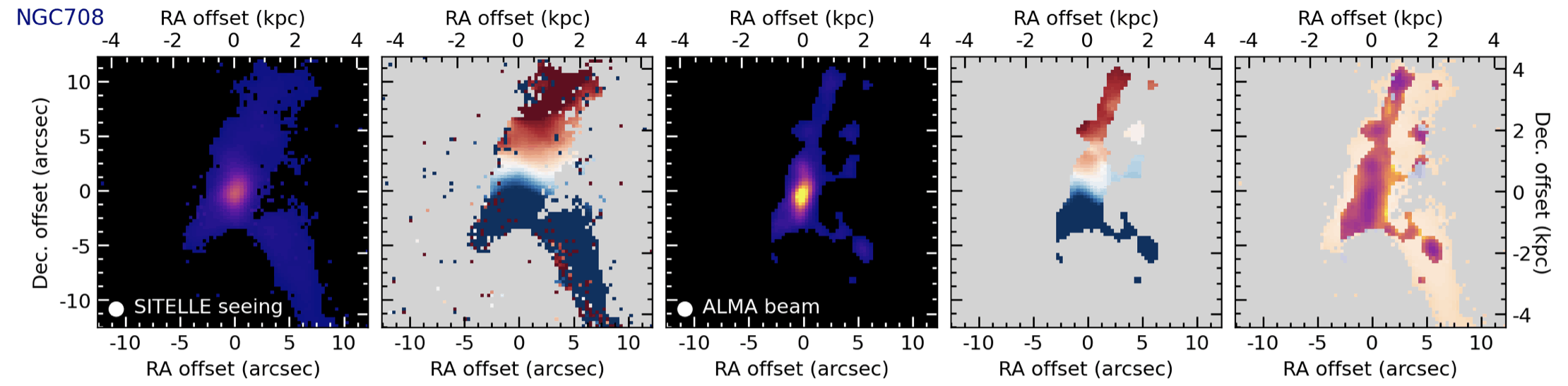}\\\includegraphics[width=0.99\textwidth]{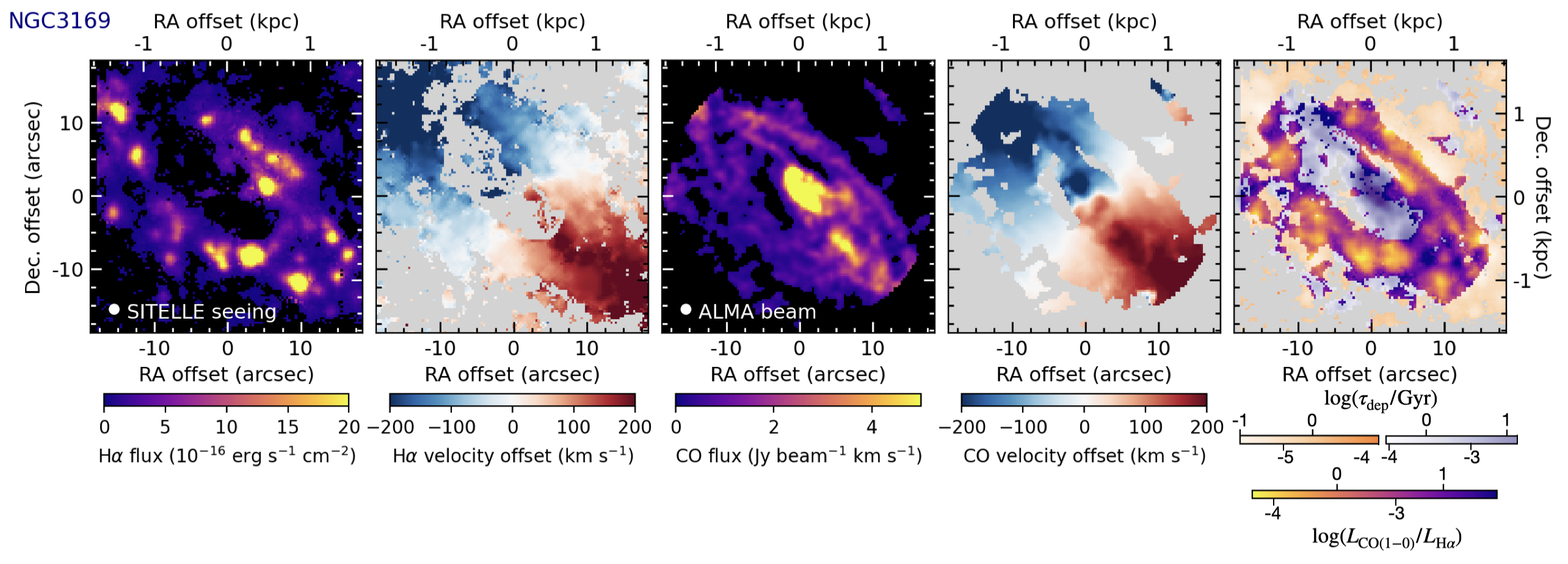}
  \caption{\label{fig:maps} Maps of NGC~383, NGC~524, NGC~708 and NGC~3169. From left to right: H$\alpha$ surface brightness map, derived from our SITELLE observations (see Section~\ref{data:Halpha} for details), H$\alpha$ luminosity-weighted mean line-of-sight velocity map (measured with respect to the systemic velocity of each galaxy listed in Table~\ref{tab1}), CO surface brightness map, derived from our ALMA observations (matching the spatial resolution of the SITELLE observations; see Section~\ref{data:CO} for details), CO luminosity-weighted mean line-of-sight velocity map (measured with respect to the systemic velocity of each galaxy listed in Table~\ref{tab1}) and CO-to-H$\alpha$ luminosity ratio ($L_{\rm CO(1-0)}$/$L_{\rm H\alpha}$) or equivalently depletion time ($\tau_{\rm dep}$) lower limit map (the orange-purple colour scale shows actual measurements, for spaxels for which both CO and H$\alpha$ are brighter than the adopted detection threshold; the light-orange colour scale shows upper limits, for spaxels for which only H$\alpha$ is brighter than the adopted detection threshold; the grey-purple colour scale shows lower limits, for spaxels for which only CO is brighter than the adopted detection threshold). For the right-most panel, we indicate $\log(L_{\rm CO(1-0)}$/$L_{\rm H\alpha})$ and $\tau_{\rm dep}$ lower limits using the lower and the upper scales of the colour tables, respectively. The $\tau_{\rm dep}$ lower limits are calculated using all H$\alpha$ emission, irrespective of ionisation mechanism. 
  }
\end{figure*}

\begin{figure*}
  \ContinuedFloat
\centering\includegraphics[width=0.98\textwidth]{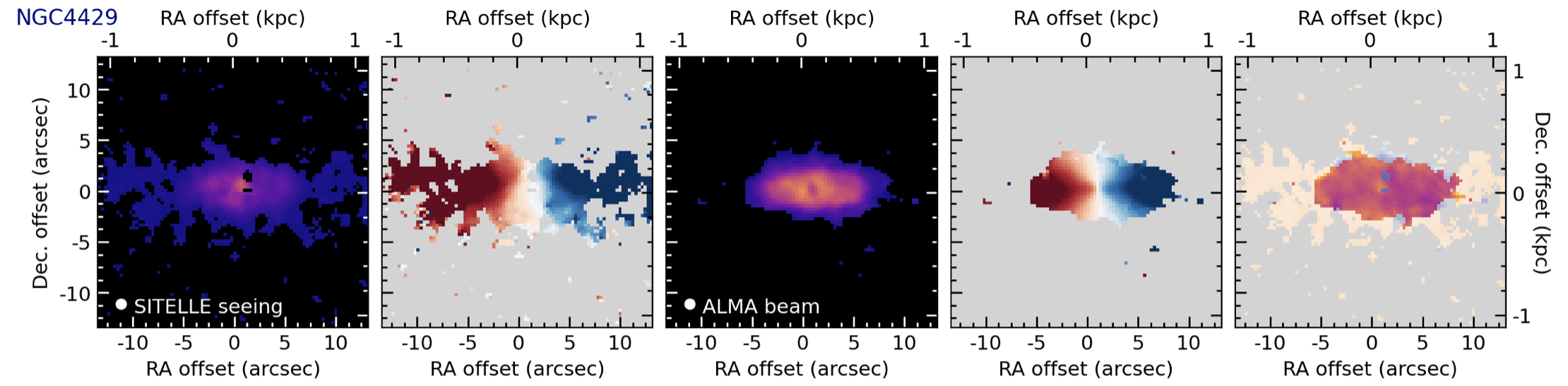}\\\includegraphics[width=0.99\textwidth]{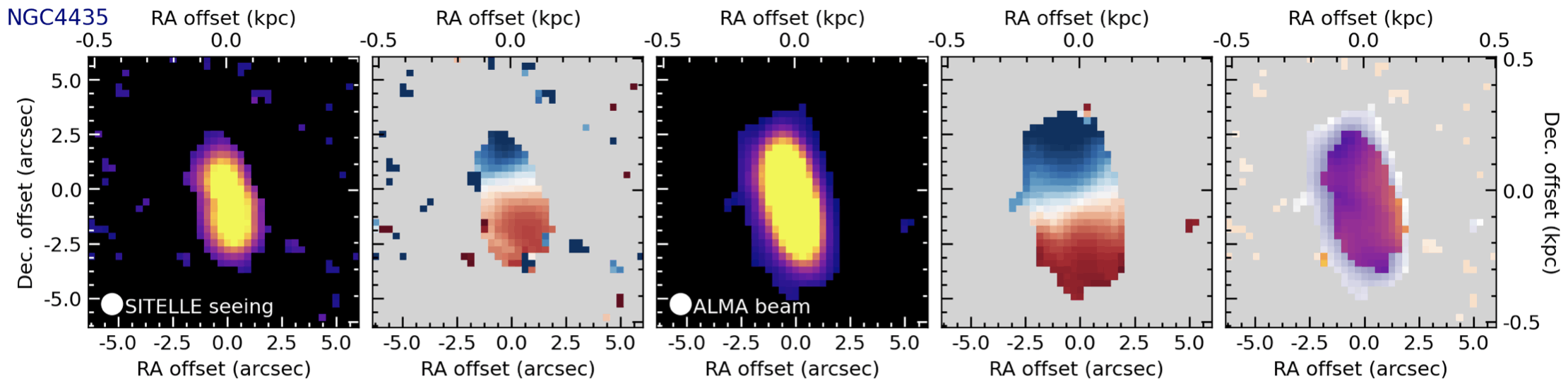}\\\includegraphics[width=0.98\textwidth]{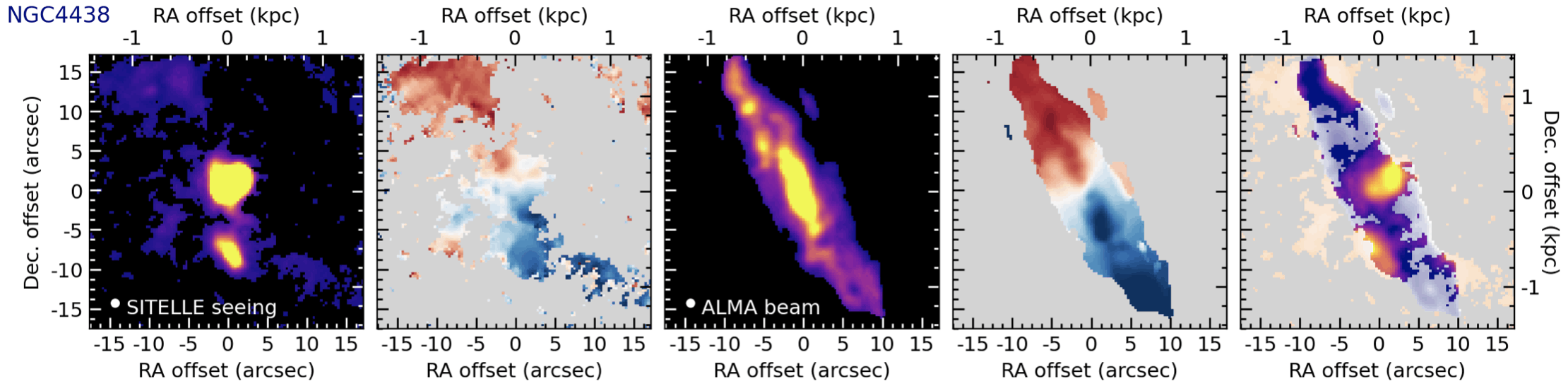}\\\includegraphics[width=0.998\textwidth]{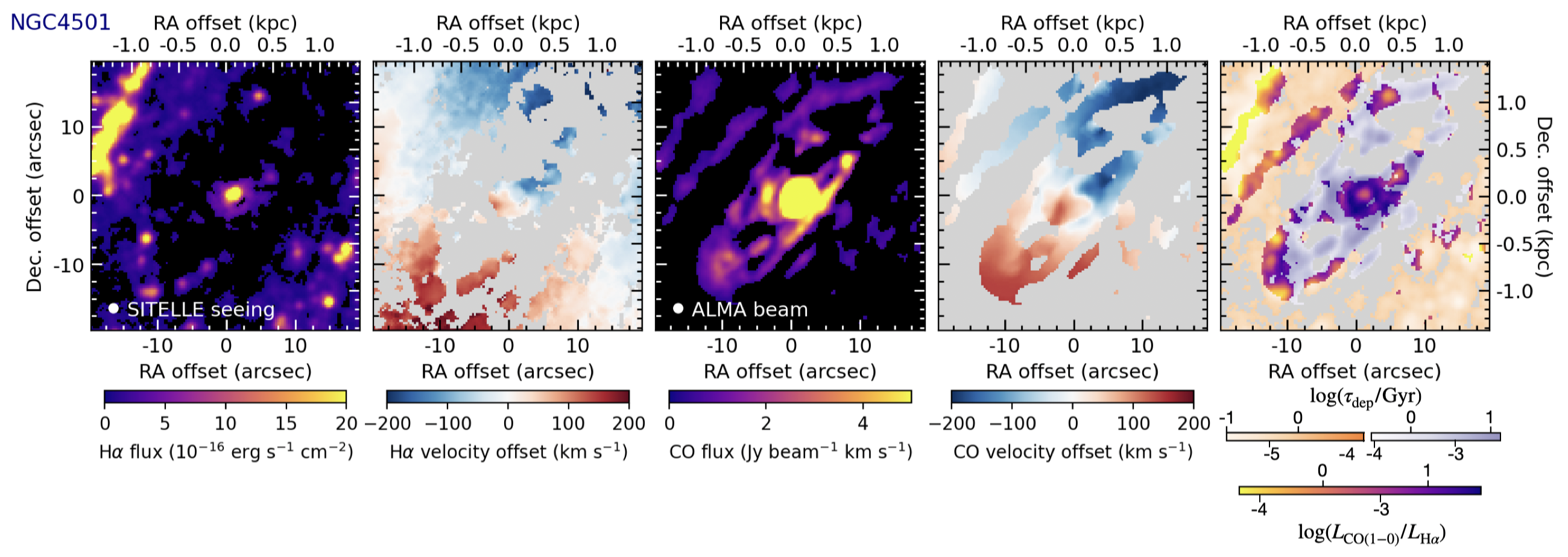}
  \caption{\label{fig:maps} (continued) Maps of NGC 4429, NGC~4435, NGC~4438 and NGC~4501.}
\end{figure*}

From the moment-0 maps, we generate molecular gas mass surface density maps. The CO flux ($F_{\rm CO}$) within each spaxel is obtained from the moment-0 map by dividing the surface brightness of each spaxel (in units of Jy~beam$^{-1}$~km~s$^{-1}$) by the synthesised beam area in spaxels. This flux is then converted into a luminosity using the following relation:
\begin{equation}
  \frac{L_{\rm CO}}{\rm K~km~s^{-1}~pc^2}=\left(\frac{3.25\times10^7}{{(1+z)}^3}\right)\,\left(\frac{F_{\rm CO}}{\rm Jy~km~s^{-1}}\right)\,{\left(\frac{\nu_{\rm obs}}{\rm GHz}\right)}^{-2}\,\left(\frac{D}{\rm Mpc}\right)^2
\end{equation}
\citep[e.g.][]{decarli2016alma}, where $z$ is the galaxy redshift, $\nu_{\rm obs}$ the observed frequency (i.e.\ the redshifted frequency of the \textsuperscript{12}CO(2-1) or \textsuperscript{12}CO(3-2) line) and $D$ the galaxy distance. The luminosity-based molecular gas mass within each spaxel is then calculated using
\begin{equation}
  \frac{M_{\rm mol}}{{\rm M}_\odot}=4.3\,\left(\frac{L_{\rm CO(1-0)}}{\rm K~km~s^{-1}~pc^2}\right)\,\left(\frac{X_{\rm CO(1-0)}}{\rm 2\times10^{20}\,cm^{-2}~(K~km~s^{-1})^{-1}}\right)\,\,\,,
\end{equation}
where $L_{\rm CO(1-0)}$ is the \textsuperscript{12}CO(1-0) luminosity and $X_{\rm CO(1-0)}$ the \textsuperscript{12}CO(1-0)-to-molecules conversion factor. We adopt a \textsuperscript{12}CO(2-1)/\textsuperscript{12}CO(1-0) ratio of $0.8$ and a \textsuperscript{12}CO(3-2)/\textsuperscript{12}CO(2-1) ratio of $1.06$ (in brightness temperature units), typical of spiral galaxies \citep[e.g.][]{lamperti2020co}. We adopt these ratios for all the galaxies in our sample, although we note that the ratio between different CO emission lines depend on the optical depth and excitation conditions of the molecular gas. For example, \citet{Ruffa2019MNRAS.484.4239R} adopted a \textsuperscript{12}CO(2-1)/\textsuperscript{12}CO(1-0) ratio of $2.3$ for radio galaxies. We adopt $X_{\rm CO(1-0)}=2.3\times10^{20}$~cm$^{-2}$~(K~km~s$^{-1}$)$^{-1}$, commonly used in extragalactic studies \citep[e.g.][]{hughes2013probability, utomo2015giant, sun2018cloud}, including the mass contribution of helium and other heavy elements \citep{strong1988radial, bolatto2013co}. Although the conversion factor can depend on the metallicity and environment of the molecular gas (e.g.\ the radiation field; \citealt{bolatto2013co}), we use a fixed $X_{\rm CO(1-0)}$ here to allow a fair comparison across our sample of ETGs and bulges (that share similar properties). The impact of these assumptions on the measured SFEs is discussed in Section~\ref{dis:Xco}. The molecular gas mass surface density within one spaxel ($\Sigma_{\rm mol}$) is then calculated as $M_{\rm mol}$ divided by the spaxel area. The uncertainty of $M_{\rm mol}$ is calculated using the RMS noise as the CO flux uncertainty in each pixel. The systematic errors on the distance measurements and the $X_{\rm CO(1-0)}$ factor are not included.

The total molecular gas mass ($M_{\rm mol,tot}$) of each galaxy measured within the ALMA FOV is listed in Table~\ref{tab1}. 
The $M_{\rm mol,tot}$ is calculated using equations (1) and (2), where $F_{\rm CO}$ is replaced with the total CO flux within the ALMA FOV (the sum of the fluxes of all the pixels within a mask defined by a signal-to-noise ratio $S/N>1.5$).
These masses are consistent with those measured independently using the same observations but different synthesised beam sizes by \citet{Davis2022MNRAS.512.1522D}.

\subsection{Ionised-gas observations}
\label{data:Halpha}

The ionised gas was observed at CFHT using SITELLE \citep{drissen2019sitelle}, an optical imaging Fourier transform spectrograph equipped with two E2V detectors each with $2048\times2064$ pixels. The SITELLE FOV is $11\arcmin\times11\arcmin$, resulting in a mean spaxel size on the sky of $\approx0\farcs31\times0\farcs31$. For each target, two data cubes were obtained: one centred on the emission lines of [\ion{N}{ii}]$\lambda6548$, H$\alpha$, [\ion{N}{ii}]$\lambda6583$, [\ion{S}{ii}]$\lambda6716$ and [\ion{S}{ii}]$\lambda6731$ with the SN3 filter ($6480$ -- $6860$~\AA) at a mean spectral resolution $R\approx2500$; the other centred on the emission lines of H$\beta$, [\ion{O}{iii}]$\lambda4959$ and [\ion{O}{iii}]$\lambda5007$ with the SN2 filter ($4840$ -- $5120$~\AA) at a mean spectral resolution $R\approx1000$. These data were taken between 2020 and 2024 as part a number of programmes: 20BC09 (PI: Boyce), 20BC25 (PI: Boyce), 22BC99 (PI: Boyce), 23AC06 (PI: Lu) and 24AC18 (PI: Lu).

The data reduction was performed with the {\tt ORBS} software developed for SITELLE \citep{martin2015orbs, martin2021data}. The seeing (full-width at half-maximum, FWHM, of the point spread function) of each target was $\approx1\arcsec$, measured from Gaussian fits to foreground stars from the Gaia catalogue \citep{lindegren2018gaia}. The SN3 data were further calibrated in wavelength based on velocity measurements of the OH sky line, allowing to use these cubes for line-of-sight velocity measurements with an absolute precision of a few km~s$^{-1}$ \citep{martin2016optimal}. Sky subtraction was performed using a median sky spectrum extracted from a $200\times200$ spaxels region located far away from each galaxy.

Our sample contains ETGs and bulges of spiral galaxies, so to measure faint emission-line fluxes, additional care must be taken to accurately remove the strong stellar continua. 
To best model the continuum of each galaxy, we tested different binning methods and modelled the stellar populations (stellar continua) using penalised pixel fitting\footnote{We used version 8.2.3 available from https://pypi.org/project/ppxf/ .} ({\tt pPXF}; \citealt{2023MNRAS.526.3273Cappellari}) and the E-MILES stellar templates \citep{2016MNRAS.463.3409Vemiles}. This process is explained in detail in Appendix~\ref{app_continuum}. We then subtract the most appropriate continuum from the original spectrum at each spaxel, resulting in a pure emission-line spectrum. After this, the emission lines are fitted using the extraction software {\tt ORCS}\footnote{https://github.com/thomasorb/orcs} \citep{2015ASPC..495..327Martin}.

For each emission line, {\tt ORCS} outputs parameters (and associated uncertainties) including the integrated flux, peak flux, intensity-weighted mean line-of-sight velocity and FWHM, and the local continuum level near the emission line. 
From these, maps of integrated flux, intensity-weighted mean line-of-sight velocity and intensity-weighted line-of-sight velocity dispersion are generated. A detection threshold is then applied to each spaxel, requiring a summed H$\alpha$ and [\ion{N}{ii}] flux with $S/N>3$. 
The H$\alpha$ surface brightness map of each galaxy is shown in Figure~\ref{fig:maps}. 
We repeated this process using the alternative data extraction software for SITELLE {\tt LUCI}\footnote{https://github.com/crhea93/LUCI} \citep{rhea2021luci}. The fluxes recovered by both algorithms agree with each other within the uncertainties. 

We correct the observed H$\alpha$ fluxes ($F_{\rm H\alpha,obs}$) for extinction using the observed H$\beta$ fluxes ($F_{\rm H\beta,obs}$) and an assumed Balmer decrement, as described below. Due to differences between the spectral resolutions and the observing conditions of the two SITELLE filters, the H$\beta$ spaxels that satisfy a $3\sigma$ detection threshold only constitute $\approx50\%$ of the H$\alpha$ spaxels satisfying that same condition. For each spaxel with both the H$\alpha$ and H$\beta$ lines satisfying the $3\sigma$ detection threshold, the "colour excess" of H$\alpha$ over H$\beta$ is defined as
\begin{equation}
  E({\rm H}\beta-{\rm H}\alpha)\equiv2.5\log\left(\frac{(F_{\rm H\alpha,obs}/F_{\rm H\beta,obs})}{({\rm H\alpha}/{\rm H\beta})_{\rm intrinsic}}\right)\,\,\,,
\end{equation}
where we assume $({\rm H\alpha}/{\rm H\beta})_{\rm intrinsic}=2.86$, as expected for case~B recombination at a temperature of $10^4$~K \citep{osterbrock2006astrophysics}. For each spaxel with a $3\sigma$ H$\alpha$ detection but no $3\sigma$ H$\beta$ detection, we adopt the H$\alpha$ extinction of the nearest reliable spaxel. 
The lack of $3\sigma$ H$\beta$ detection in many spaxels is likely due to high continua hindering the detection of H$\beta$ emission. Based on the optical images, our targets do not seem to be dusty enough to completely block H$\beta$ emission. 

The H$\alpha$ extinction is then calculated as
\begin{equation}
  A_{\rm H\alpha}=\left(\frac{E({\rm H\beta}-{\rm H\alpha})}{k(\lambda_{\rm H\beta})-k(\lambda_{\rm H\alpha})}\right)\,k(\lambda_{\rm H\alpha})
\end{equation}
following \citet{nelson2016spatially}, where $k(\lambda)$ is the reddening curve of \citet{fitzpatrick1986average} and $k(\lambda_{\rm H\alpha})$ and $k(\lambda_{\rm H\beta})$ are evaluated at the wavelengths of H$\alpha$ and H$\beta$, respectively. 
For each galaxy, we also calculate the integrated $V$-band extinction $A_V$ ($A_V=3.1\,E{\rm (B-V)}$, where $E{\rm (B-V)}=E{\rm(H\beta-H\alpha)}/1.07$), from the ratio of the sum of the H$\alpha$ and the H$\beta$ fluxes in spaxels where both H$\alpha$ and H$\beta$ satisfy the $3\sigma$ detection threshold, listed in Table~\ref{tab1}.
Finally, the extinction-corrected H$\alpha$ flux ($F_{\rm H\alpha}$) is calculated as
\begin{equation}
  F_{\rm H\alpha}=F_{\rm H\alpha,obs}\,e^{A_{\rm H\alpha}/1.086}\,\,\,.
\end{equation}

This extinction-corrected H$\alpha$ flux is converted to a SFR using the relation of \citet{kennicutt2012star}:
\begin{equation}
  \log({\rm SFR}\,/\,{\rm M}_\odot\,{\rm yr}^{-1})=\log(L_{\rm H\alpha}\,/\,{\rm erg}\,{\rm s}^{-1})-41.27\,\,\,,
\end{equation}
where $L_{\rm H\alpha}=F_{\rm H\alpha}(4\pi D^2)$ is the extinction-corrected H$\alpha$ luminosity. We note that when probing spatial scales smaller than $\approx500$~pc, this conversion relation can break down, as seen e.g.\ in \citet{kennicutt2012star}. Local SFRs depend on the environment and age of the stellar population, which we do not consider here. Nevertheless, this conversion holds true for radial profiles of the depletion time (see Section~\ref{dep_rad}), that are calculated within apertures sufficiently large for robust H$\alpha$-to-SFR conversions. 

The surface density of SFR ($\Sigma_{\rm SFR}$) within one spaxel is calculated as the SFR within that spaxel divided by the spaxel area. 
The total SFR of each target is calculated using equations (5) and (6), where $F_{\rm H\alpha,obs}$ is the sum of the H$\alpha$ fluxes within the galaxy and $A_{\rm H\alpha}$ is taken as the mean of the extinctions in spaxels with $3\sigma$ H$\alpha$ and H$\beta$ detections. The uncertainty of this total SFR is calculated from the uncertainties of the H$\alpha$ and H$\beta$ fluxes in each spaxel used to calculate it. 
The total SFR of each target and its uncertainty are listed in Table~\ref{tab1}. We compared those SFRs to those reported by \citet{Davis2022MNRAS.512.1522D}, measured from integrated {\it GALEX} FUV and {\it WISE} $22$-$\mu$m flux densities. Considering the uncertainties of our flux measurements, that can be as high as $20$\%, the two SFR measurements generally agree with each other.

Having said that, we note that the SFRs calculated here are really upper limits, as large fractions of the H$\alpha$ fluxes do not originate from star-forming regions. This is confirmed by the ionisation mechanism classifications (see Section~\ref{BPT}), suggesting that the majority of the observed ionised-gas emission does not stem from ionisation by SF. $22$-$\mu$m emission may trace dust heated by old stellar populations and the UV-upturn phenomenom has been observed in the associated low-ionisation emission-line regions (LIERs; see e.g.\ \citealt{2011ApJS..195...22Yi, 2019AJ....158....2Byler}). However, he SFRs traced by FUV and $22$-$\mu$m, shown in Table~\ref{tab1}, had the contribution from old stellar populations  subtracted, and are thus more accurate representation of the true SFRs.

\subsection{CO-to-H$\alpha$ luminosity ratio}
\label{dep}

In the right-most panels of Figure~\ref{fig:maps}, we show the ratios of the CO and the H$\alpha$ luminosities ($L_{\rm CO(1-0)}/L_{\rm H\alpha}$), based on the calculations detailed in Sections~\ref{data:CO} and \ref{data:Halpha}. 
For spaxels for which both $L_{\rm CO(1-0)}$ and $L_{\rm H\alpha}$ are reliably measured (i.e.\ for which both the H$\alpha$ line and the CO line are detected with $S/N\ge3$), $L_{\rm CO(1-0)}/L_{\rm H\alpha}$ is calculated by simply dividing the two luminosities. This ratio is shown in Figure~\ref{fig:maps} with an orange-purple colour scale. However, many spaxels have only one tracer with a $S/N$ higher than our adopted detection threshold. When only $L_{\rm H\alpha}$ is reliably measured, we estimate a $L_{\rm CO(1-0)}/L_{\rm H\alpha}$ upper limit by dividing the $3\sigma$ upper limit of $L_{\rm CO(1-0)}$ by the reliably measured $L_{\rm H\alpha}$, shown in Figure~\ref{fig:maps} with a light-orange colour scale. An analogous approach is applied to spaxels for which only $L_{\rm CO(1-0)}$ is reliably measured, to estimate a $L_{\rm CO(1-0)}/L_{\rm H\alpha}$ lower limit, shown in Figure~\ref{fig:maps} with a grey-purple colour scale.

The $L_{\rm CO(1-0)}/L_{\rm H\alpha}$ maps in Figure~\ref{fig:maps} show that different galaxy types are distinct. Since the CO and the H$\alpha$ emission lines trace the molecular and the ionised gas, respectively, we use these maps to trace the different phases of the gas in our sample galaxies.
ETGs tend to have smooth and overlapping molecular and ionised-gas discs. This indicates that the cold molecular gas and the warm ionised-gas phases in these ETGs are tightly correlated, both spatially and in terms of their kinematics. In particular, NGC~708 is a brightest cluster galaxy (BCG) where the cold gas has been suggested to form via cooling of the halo hot gas \citep{2021MNRAS.503.5179North}. The fact that the two gas phases observed here are tightly correlated and share the same filamentary distribution strongly supports this hypothesis. This will be further discussed in Section~\ref{dis:AGN}.  The three spiral galaxies in our sample, however, have spatially decorrelated peaks of molecular and ionised gas. 
NGC~4501 and NGC~3169 have star-forming rings just outside the bulges. This is most clearly illustrated by the $L_{\rm CO(1-0)}/L_{\rm H\alpha}$ map of NGC~3169. As for NGC~4501, the $L_{\rm CO(1-0)}/L_{\rm H\alpha}$ map of its star-forming ring is less informative because the molecular gas is almost completely depleted at the location of H$\alpha$ emission peaks, leaving very few overlapping regions of CO(1-0) and H$\alpha$ emission. NGC~4438 has a different morphology, with bright ionised-gas emission and high $\Sigma_{\rm mol}$ at its centre. This is likely due to its interaction with NGC~4435 and the induced morphological distortion and nuclear activity.

If we convert the H$\alpha$ luminosities to SFR upper limits and the CO luminosities to molecular gas masses, the $L_{\rm CO(1-0)}/L_{\rm H\alpha}$ maps can be interpreted as maps of depletion time lower limits. In later sections, we show the spatially-resolved classification of the ionisation mechanisms and discuss the uncertainties of the SFR and molecular gas mass conversions.

\section{Results}
\label{results}

\subsection{Ionised-gas line ratios}
\label{BPT}

As stated in previous sections, the ETGs and bulges in our sample all have old stellar populations and each hosts an AGN or a compact radio source in its nucleus (summarised in Table~\ref{tab2}). Therefore, in each case it is necessary to disentangle the fraction of the ionised-gas emission that arises from regions dominated by SF from that contaminated by significant emission from other processes. In our SITELLE datacubes, we have a suite of ionised-gas emission lines that can be used to distinguish different ionisation mechanisms and identify SF regions. We adopt two methods: the Baldwin, Phillips \& Terlevich (BPT) classification (as described in \citealt{1981PASP...93....5BPT}) and the H$\alpha$ equivalent width ($W_{\rm H\alpha}$) versus [\ion{N}{ii}]/H$\alpha$ (WHAN) classification \citep{2010MNRAS.403.1036Cid, 2011MNRAS.413.1687Cid}.

\begin{table*}
\caption{Nuclear properties of the sample galaxies.}
\label{tab2}
\centering
\begin{tabular}{|l|l|l|l|l|l|} \hline  
Target & Nuclear activity & AGN type & $M_{\rm BH}$ & Method & References\\  
 & & & (M$_\odot$)& &\\  
 (1)& (2)& (3)& (4)& (5)&(6)\\ \hline 
NGC~\phantom{0}383 & Jetted radio-loud AGN & LERG & $4.2\phantom{4}\times10^9$ & Gas dynamics & \citet{2002MNRAS.336.1161Laing}, \citet{2019MNRAS.490..319North}\\   
NGC~\phantom{0}524 & Compact radio source& - & $4.0\phantom{4}\times10^8$ & Gas dynamics & \citet{Nyland2016MNRAS.458.2221N}, \citet{Smith2019MNRAS.485.4359S}\\   
NGC~\phantom{0}708 & Brightest cluster galaxy & LERG & $2.9\phantom{4}\times10^8$ & Stellar velocity dispersion & \citet{2002ApJ...579..530Woo}, \citet{2021MNRAS.503.5179North}\\ 
NGC~3169 & Low-luminosity AGN & Seyfert~1 & $1.6\phantom{4}\times10^8$ & Bulge mass & \citet{nagar2005radio}, \citet{2006AJ....131.1236Dong}\\   
NGC~4429 & Low-luminosity AGN & - & $1.5\phantom{4}\times10^8$ & Gas dynamics & \citet{Nyland2016MNRAS.458.2221N}, \citet{2018MNRAS.473.3818Davis}\\   
NGC~4435$^\dagger$ & No AGN & - & $7.5\phantom{4}\times10^6$ & Gas dynamics & \citet{2006MNRAS.366.1050Coccato}, \citet{2007ApJ...656..206Panuzzo}\\   
NGC~4438$^\dagger$ & Double-lobed radio-loud AGN & LINER & $5.0\phantom{4}\times10^7$ & Bulge mass & \citet{2004ApJ...610..183Machacek}, \citet{2007MNRAS.380.1009Hota}\\   
NGC~4501 & Radio-quiet AGN & Seyfert~2 & $1.34\times10^7$ & Stellar dynamics & \citet{2010ASPC..427..377Panessa}, \citet{2017MNRAS.471.2187Davis}\\ 
\hline 
\end{tabular}
{\it Notes:} {(1) Galaxy name. The interacting galaxy pair, NGC~4438 and NGC~4435, is marked with daggers. (2) Description of nuclear activity, with references listed in column (6). (3) AGN type or high-/low-excitation radio galaxy (HERG/LERG) classification taken from \citet{2024MNRAS.528..319Elford}. (4) SMBH mass, with references listed in column (6). (5) Method used to measure the SMBH mass. (6) Relevant studies of nuclear activity and SMBH mass.}
\end{table*}

\begin{figure*}
  \centering\includegraphics[width=0.95\textwidth]{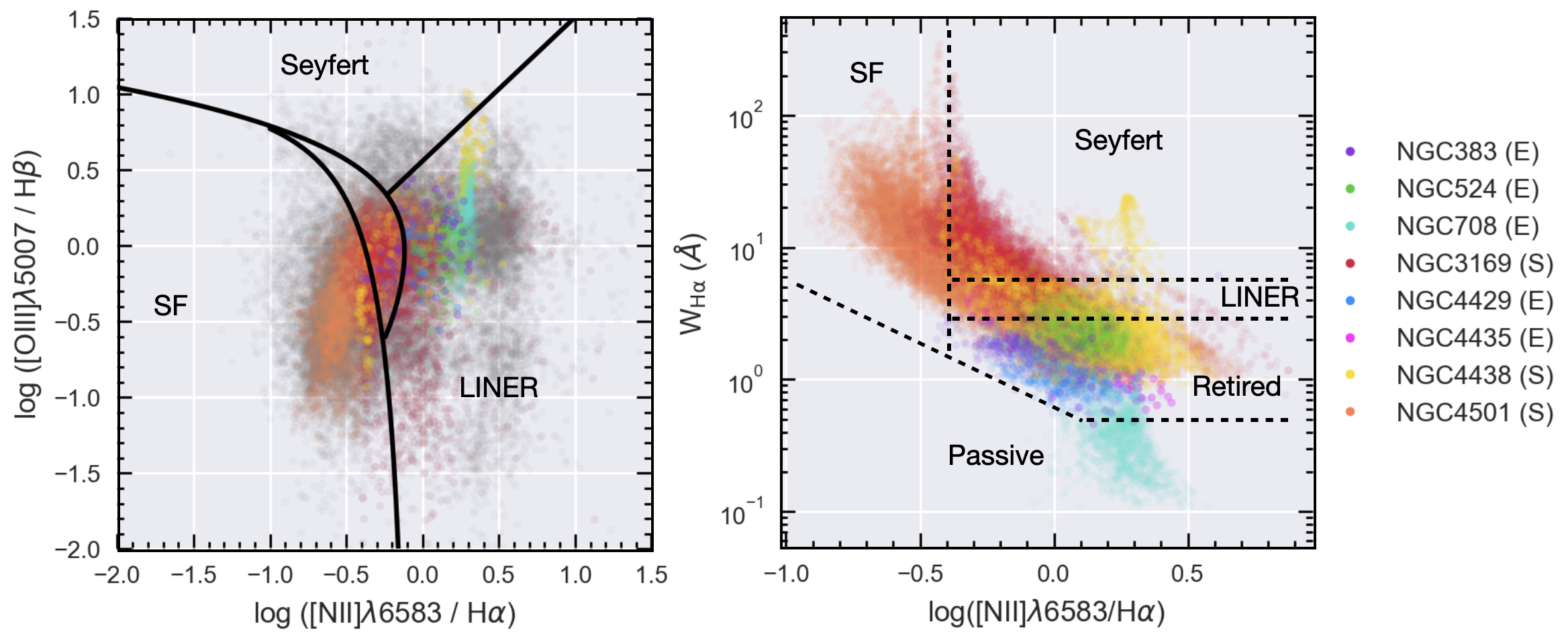}\\\includegraphics[width=0.95\textwidth]{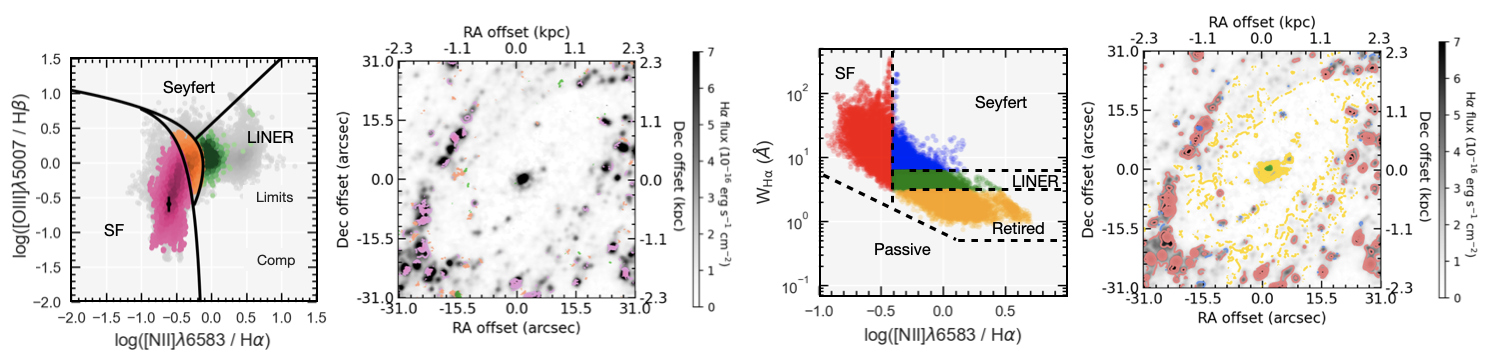}
  \caption{\label{fig:BPT} {\bf Top row:} ionisation-mechanism classification using emission-line ratios. Left: BPT diagram, classifying the ionisation based on the line ratios [\ion{N}{ii}]/H$\alpha$ and [\ion{O}{iii}]/H$\beta$ \citep{1981PASP...93....5BPT}. Data points from the spaxels of each galaxy with all four emission-line integrated fluxes having $S/N\ge3$ are colour coded. Grey data points indicate spaxels for which one of the emission lines in either [\ion{N}{ii}]/H$\alpha$ or [\ion{O}{iii}]/H$\beta$ has an integrated flux with $S/N<3$, and hence are only limits. The ionisation classification boundaries are taken from \citet{kewley2006host} and are marked by the black solid lines. Right: WHAN diagram, classifying the ionisation based on the line ratio [\ion{N}{ii}]/H$\alpha$ and the equivalent width of H$\alpha$. Data points from the spaxels of each galaxy with the two emission-line integrated fluxes having $S/N\ge3$ are colour coded. The ionisation classification boundaries are taken from \citet{2011MNRAS.413.1687Cid} and are marked by the black dashed lines. Using these two distinctive classifications reveals that our sample ETGs do not have SF nor strong AGN ionisation, but that the gas is instead likely ionised by old stellar populations. The BPT and WHAN diagrams of each individual galaxy, along with the locations of the different ionisation mechanisms, are shown in Appendix~\ref{app_lineratios}.
  Bottom row: example of the application of the two classification methods to the galaxy NGC~4501. In the BPT (left panel) and WHAN (middle-right panel) diagrams, the data are colour-coded according to the dominant ionisation source. The locations of the different ionisation sources are overlaid on the H$\alpha$ integrated flux maps using matching colours (middle-left and right panels).
  }
\end{figure*}

The BPT diagram uses the [\ion{O}{iii}]/H$\beta$ and [\ion{N}{ii}]/H$\alpha$ ratios ("O3N2" type) or [\ion{O}{iii}]/H$\beta$ and [\ion{S}{ii}]/H$\alpha$ ratios ("O3S2" type) to separate galaxies into SF, Seyfert (strong AGN activity), low-ionisation nuclear emission region (LINER; weak AGN activity and/or old stellar populations) and composite (a mix of AGN, old stellar populations and SF activity). Due to the low $S/N$ of the [\ion{S}{ii}] emission lines in most of our SITELLE data cubes, we use the "O3N2" type of BPT diagram to classify our sample galaxies. We extract the emission line fluxes of H$\alpha$, [\ion{N}{ii}], H$\beta$ and [\ion{O}{iii}] from the spectrum at each spaxel and adopt the classification boundaries of \citet{kewley2006host} to separate and uniquely identify SF-ionised, Seyfert/LINER-ionised and composite regions. When all four of the emission lines have integrated fluxes with $S/N\ge3$, we consider the line ratios well constrained. These data are shown as coloured data points in the left panel of Figure~\ref{fig:BPT}. When one of the emission lines has an integrated flux below this $S/N$ threshold, we use the continuum level to estimate an upper limit of the integrated flux. These data are shown as grey data points in the left panel of Figure~\ref{fig:BPT}. We also identify the locations of these different ionisation mechanisms within each galaxy, as shown in Appendix~\ref{app_lineratios}.

Based on the BPT diagram, only the three spiral galaxies have reliably-identified regions where SF dominates the ionisation, whose locations are shown in Appendix~\ref{app_lineratios}. As expected, in NGC~4501 and NGC~3169 these SF regions reside in the star-forming rings. In NGC~4438 the SF regions are solely present in the second brightest emission-line region near the galaxy centre.  
The majority of the spaxels in our ETGs are classified as LINER-ionised. However, most of these spaxels have very weak H$\beta$ and [\ion{O}{iii}] emission lines, contaminated by high continua and strong absorption features, as illustrated by the SN2 spectra shown in Appendix~\ref{app_continuum}. The BPT diagram therefore does not provide a robust classification of the ionisation mechanisms in these galaxies.

The WHAN diagram \citep{2010MNRAS.403.1036Cid} relies on the [\ion{N}{ii}] and H$\alpha$ emission only, and is hence useful to classify low $S/N$ emission-line regions. The emission-line regions are classified into five categories: 1) SF: star formation; 2) Seyfert: strong AGN activity; 3) LINER: weak AGN activity; 4) retired: old stellar populations; and 5) passive: very weak or undetected emission lines. The WHAN diagram has been successfully applied to Sloan Digital Sky Survey observations of galaxies \citep{2011MNRAS.413.1687Cid} and several kpc-scale studies \citep[e.g.][]{2020AAS...23525802Greene, 2024MNRAS.528.5252Mezca}. However, its application to $100$~pc-scale data is novel. Here, we therefore verify whether the kpc-scale WHAN classification boundaries can be used at $100$~pc scale, by comparing the outcomes with the BPT classifications.

We extract the [\ion{N}{ii}] fluxes, H$\alpha$ fluxes and $W_{\rm H\alpha}$ from the SITELLE spectra, and adopt the classification of \citet{2011MNRAS.413.1687Cid}. As shown in the right panel of Figure~\ref{fig:BPT}, SF regions are again present exclusively in the three spiral galaxies of our sample. We show the locations of these SF regions in Appendix~\ref{app_lineratios}, along with those of regions ionised by other mechanisms. The locations of the SF regions agree well with those classified using the BPT diagram. 

The high $S/N$ of the H$\alpha$ and [\ion{N}{ii}] integrated fluxes however allow to include many more spaxels in the WHAN diagram than in the BPT diagram, hence we are able to classify most of the regions in the bulges of the spiral galaxies and ETGs. 
Most of the spaxels in the ETGs are classified as "retired", implying that the emission arises from old stellar populations \citep{2011MNRAS.413.1687Cid}. The centres of some of our targets have Seyfert ionisation, surrounded by LINER ionisation (however not confined to the nuclear regions, therefore LIER ionisation), indicating that AGN ionisation is typically confined to the central $\approx200$~pc in radius. Our results thus show that, when present, an AGN can impact the ionisation level of its surroundings, but its effect is negligible on galactic scales. This is not surprising considering that each sample galaxy either has a low-luminosity AGN or is classified as a low-excitation radio galaxy (LERG). LERGs (NGC~383 and NGC~708 in our sample) have weak or absent low-ionisation optical emission lines and host radio-mode AGN, so the dominant energy output from the AGN is in the form of radio jets that do not produce radiation sufficient to strongly affect the surrounding gas discs (see e.g.\ \citealt {2020MNRAS.499.5719Ruffa} and \citealt{2021MNRAS.503.5179North} for radio studies of NGC~383 and NGC~708). 
Indeed, the ionised-gas emission at every spaxel in NGC~708 is classified as "retired" or "passive", implying that NGC~708 has a strong stellar continuum and weak emission lines, mostly arising from old stars.

We now compare the BPT and WHAN diagram classifications, revealing their similarities and discrepancies. As an example (see the bottom row of Figure~\ref{fig:BPT}), the WHAN diagram of NGC~4501 captures all the SF regions in the spiral arms, that are consistent with the BPT classifications. However, there are few reliably measured H$\beta$ and [\ion{O}{iii}] fluxes within those same spiral arms, so use of the BPT diagram is much more limited. The WHAN diagram also helps to identify LINER/LIER ionisation at the very centre and old stellar populations in the outskirts of the bulge. This agrees well with a visual inspection of the galaxy and previous studies \citep[e.g.][]{moreno2016using, brum2017dusty, repetto20172d}. The WHAN diagram also suggests Seyfert ionisation in the spiral arms, indicating that these emission-line regions have particularly high $W_{\rm H\alpha}$. This is likely associated with diffuse ionised gas leaking from \ion{H}{ii} regions \citep{2022A&A...659A..26Belfiore}. 

Our work therefore not only shows the strengths of the WHAN diagram, but it also offers insight into the physics behind the ionisation mechanisms. We can therefore now begin to piece together the processes setting the emission-line ratios and equivalent widths, arising from physics at spatial scales $\sim100$~pc.

\subsection{$\Sigma_{\rm SFR}-\Sigma_{\rm mol}$ relation}
\label{KS}

\begin{figure*}
\centering\includegraphics[width=0.98\textwidth]{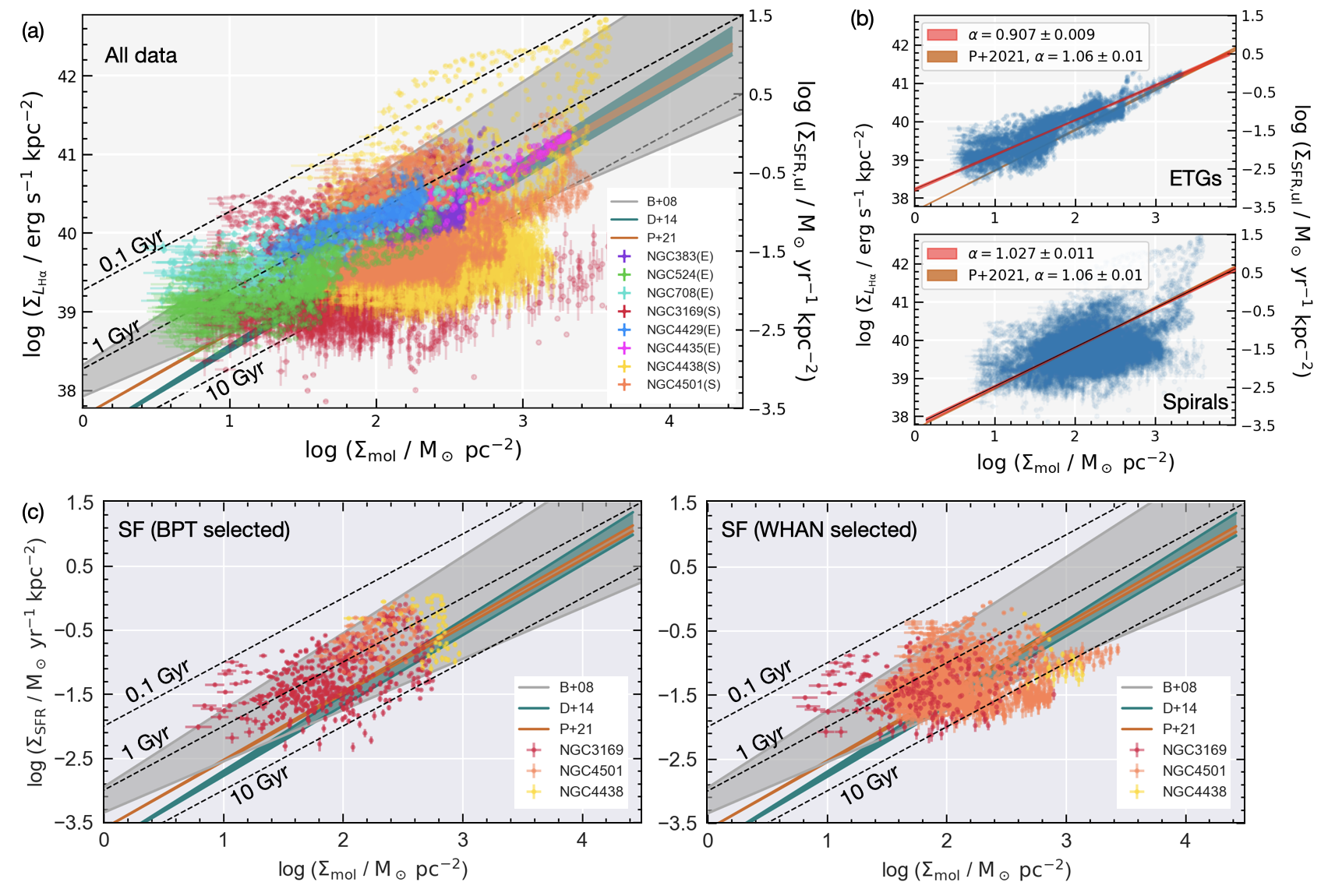}
\caption{\label{fig:dept_KS} The $\Sigma_{L_{\rm H\alpha}}$($\Sigma_{\rm SFR}$) -- $\Sigma_{\rm mol}$ relations of the individual spaxels of our sample ETGs and bulges. (a)-(b) All H$\alpha$ emission (converted to SFRs), irrespective of the ionisation mechanism. The measured (extinction-corrected) H$\alpha$ luminosity surface density of each spaxel is shown on the left-hand ordinate; the inferred SFR upper limit on the right-hand ordinate. Coloured data points, shaded by their density, have both extinction-corrected H$\alpha$ and CO integrated fluxes with $S/N\ge3$. The grey, teal and brown regions show the power-law relations (and their scatters) of \citet{bigiel2008star}, \citet{Davis2014MNRAS.444.3427D} and \citet{2021A&A...650A.134Pessa}, respectively. The black dashed diagonal lines are lines of equal depletion times (labelled). (b) All the data points separated into ETGs (top panel) and spiral galaxies (bottom panel). The best-fitting power-law relation of each category with its $98\%$ confidence interval is shown in red, with its slope listed in the legend. The power-law relation of \citet{2021A&A...650A.134Pessa} is shown in brown for comparison. The $\Sigma_{L_{\rm H\alpha}}$($\Sigma_{\rm SFR}$) -- $\Sigma_{\rm mol}$ relation of our ETGs has less scatter and a shallower slope than those of the \citet{2021A&A...650A.134Pessa} relation. 
(c) As (a), but for reliably-identified SF regions only, as identified using the BPT (left) and the WHAN (right) diagram. Only the three spiral galaxies have reliably-identified SF regions; these have $\Sigma_{\rm SFR}$ -- $\Sigma_{\rm mol}$ relations similar to that of the SF regions of nearby spiral galaxies.
  }
\end{figure*}

The $\Sigma_{\rm SFR}$ -- $\Sigma_{\rm mol}$ relation is a useful tool to probe the depletion times of different galaxies. Figure~\ref{fig:dept_KS} shows the $\Sigma_{\rm SFR}$ -- $\Sigma_{\rm mol}$ relation of all the spaxels of our sample galaxies with reliably-detected H$\alpha$ and CO (integrated fluxes with $S/N\ge3$). These datapoints are not all independent, as one resolution element is spread across approximately $10$ spaxels. The distribution of data points is compared to the characteristic timescales ($0.1$, $1$ and $10$~Gyr) and power-law relations of \citeauthor{bigiel2008star} (\citeyear{bigiel2008star}; grey region) for LTGs on kpc scales, \citeauthor{Davis2014MNRAS.444.3427D} (\citeyear{Davis2014MNRAS.444.3427D}; teal region) for ETGs and \citeauthor{2021A&A...650A.134Pessa} (\citeyear{2021A&A...650A.134Pessa}; brown region) for LTGs on $100$~pc scales.
In panels (a) and (b), we include all H$\alpha$ data, converted to SFRs irrespective of the ionisation mechanism. In panel (b), we show the ETGs and spiral galaxies separately. 
The best-fitting power-law relation of each category and its $98\%$ confidence interval is shown in red. For comparison, the power-law relation of \citeauthor{2021A&A...650A.134Pessa} (\citeyear{2021A&A...650A.134Pessa}; $100$~pc-scale studies of nearby LTGs) is shown in brown.

We recall however that the SFRs shown in these panels are truly upper limits, as indicated by the axis labels ($\Sigma_{\rm SFR,ul}$). For clarity, we therefore show the measured (extinction-corrected) H$\alpha$ luminosity surface densities ($\Sigma_{L_{\rm H\alpha}}$) on the left-hand ordinate and the inferred SFR upper limits on the right-hand ordinate. This way, the figures can also be interpreted purely as empirical correlations between molecular gas mass and H$\alpha$ luminosity surface densities.

The $\Sigma_{L_{\rm H\alpha}}$ ($\Sigma_{\rm SFR,ul}$) and $\Sigma_{\rm mol}$ of our ETGs are more tightly correlated (scatter about the best-fitting relation $\sigma=0.27$~dex) than those of our spiral galaxies  ($\sigma=0.61$~dex) and those of the LTGs studied by \citeauthor{2021A&A...650A.134Pessa} (\citeyear{2021A&A...650A.134Pessa}; $\sigma=0.41$~dex). 
As demonstrated in Section~\ref{BPT}, none of the spaxels of the ETGs has its ionisation dominated by SF. Therefore, we cannot accurately measure SFRs in our ETGs using solely our ionised-gas observations. The inferred $\tau_{\rm dep}$ are thus merely lower limits. Conversely, the inferred SFEs are upper limits. Nevertheless, the $\Sigma_{L_{\rm H\alpha}}$ -- $\Sigma_{\rm mol}$ relations of our ETGs indicate that the molecular and the ionised-gas phases of our sample ETGs are tightly correlated. 
Of course, the BPT and WHAN classifications (see Section~\ref{BPT}) show that the majority of the emission-line regions of our ETGs and bulges are classified as LINER/LIER. The H$\alpha$ emission is thus likely tracing old stellar populations, or more specifically the hot but low-mass evolved stars \citep{2012IAUS..284..132Flores}. 
It is tempting to interpret our observations as a tight empirical correlation between molecular gas and stellar mass, although previous works have shown that molecular gas and stellar luminosity are not correlated in ETGs \citep[see e.g.][]{Young2011MNRAS.414..940Y}. 

The three spiral galaxies in our sample have very scattered $\Sigma_{L_{\rm H\alpha}}$ -- $\Sigma_{\rm mol}$ relations. The large scatters can be explained by a mixture of ionisation mechanisms. NGC~4438 has the largest scatter, potentially due to its interaction with a companion, triggering multiple ionisation mechanisms including shocks \citep{2009A&A...496..669Vollmer}.

In panel (c) of Figure~\ref{fig:dept_KS}, only regions classified as SF-dominated are shown, using respectively the BPT (left) and the WHAN (right) diagram. These are thus true $\Sigma_{\rm SFR}$ -- $\Sigma_{\rm mol}$ relations (rather than $\Sigma_{\rm SFR,ul}$ -- $\Sigma_{\rm mol}$ relations). The star-forming spaxels are all from the three spiral galaxies of our sample. The very few SF spaxels identified using the BPT diagram are roughly contained within the scatter of the \citet{bigiel2008star} power-law relation, while the SF regions identified using the WHAN diagram occupy a broader range of depletion times. This is the result of more emission-line regions (spatially decorrelated with molecular clouds) being included in the WHAN diagram. For NGC~4501, many fewer spaxels are classified as SF using the BPT diagram, primarily due to the fact that the BPT diagram relies on the [\ion{O}{iii}] and H$\beta$ emission lines that are particularly faint (and thus do not meet the $S/N$ cut) in this galaxy. The larger scatters of the $\Sigma_{\rm SFR}$ -- $\Sigma_{\rm mol}$ relations of our spiral galaxies compared to those of our ETGs are potentially due to the fact that the ionised gas and molecular gas peaks are generally not co-spatial. It is thus common to have broader $\Sigma_{\rm SFR}$ -- $\Sigma_{\rm mol}$ distributions at $100$~pc scales than kpc scales \citep[e.g.][]{2010ApJ...722.1699Schruba,2021A&A...650A.134Pessa}. In fact, these scatters can be interpreted as a sign of rapid cycling between gas and stars on sub-kpc scales, in turn allowing us to constrain the molecular gas lifetimes, as shown successfully for NGC~3169 by \citet{2022MNRAS.514.5035Lu} and for other nearby galaxies by e.g.\ \citet{2019Natur.569..519Kruijssen}, \citet{2020MNRAS.496.2155Zabel} and \citet{chevance2022life}.

\subsection{Depletion time radial profiles}
\label{dep_rad}

\begin{figure*}
\centering\includegraphics[width=0.95\textwidth]{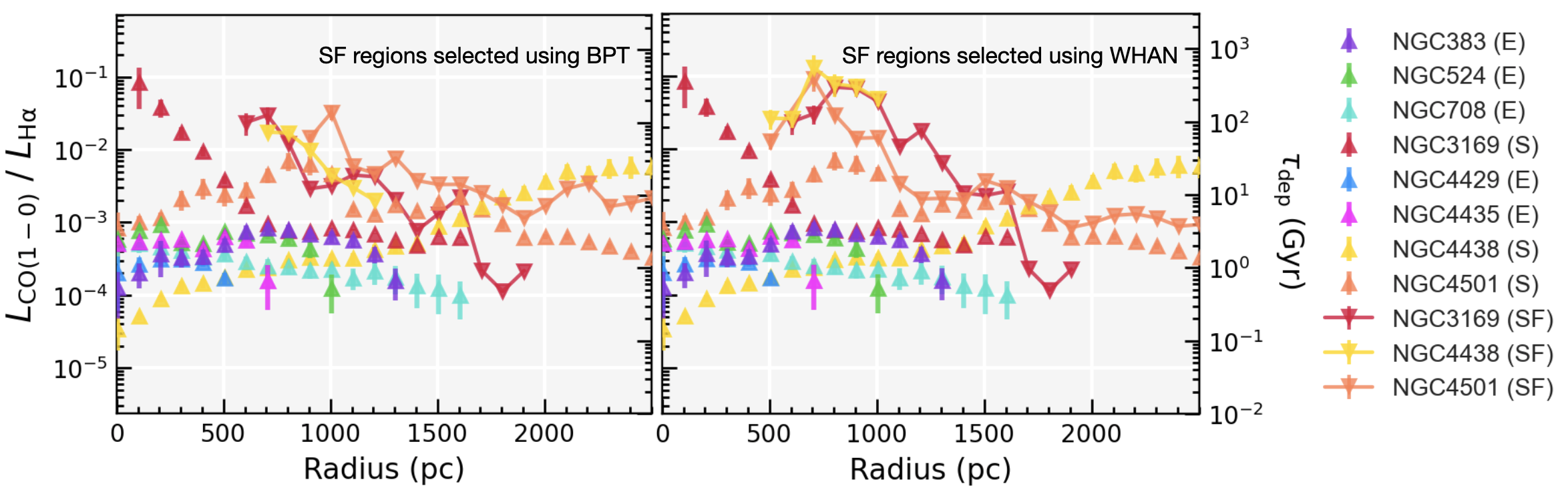}\caption{\label{fig:dept_rad} $L_{\rm CO(1-0)}$/$L_{\rm H\alpha}$ (i.e.\ depletion times) as a function of galactocentric distance (see Section~\ref{dep_rad} for details). 
The $L_{\rm CO(1-0)}$/$L_{\rm H\alpha}$ ratios are measured within annuli of $200$~pc width centred on each galaxy centre, with a step size of $100$~pc, by taking the ratio of the sum of $L_{\rm CO(1-0)}$ and the sum of $L_{\rm H\alpha}$ within each annulus. The depletion times are calculated by converting the $L_{\rm CO(1-0)}$ and $L_{\rm H\alpha}$ to molecular gas masses and SFRs, respectively. 
  The up-pointing triangles indicate $\tau_{\rm dep}$ calculated using all the H$\alpha$ emission, irrespective of the ionisation mechanism, and are thus lower limits. The down-pointing triangles indicate $\tau_{\rm dep}$ calculated using only the H$\alpha$ emission reliably-classified as star forming, and are thus upper limits. Left: SF regions selected using the BPT diagram (see Section~\ref{BPT}). Right: SF regions selected using the WHAN diagram (see Section~\ref{BPT}). Except for NGC~383, the $\tau_{\rm dep}$ lower limits of our ETGs increase with decreasing radius. The $\tau_{\rm dep}$ upper limits of our spirals also increase with decreasing radius. There is no SF region in any ETG nor within $500$~pc in radius in any spiral galaxy. 
  }
\end{figure*}

Figure~\ref{fig:dept_rad} shows the $L_{\rm CO(1-0)}/L_{\rm H\alpha}$ ratio as a function of deprojected galactocentric radius. This can be interpreted purely as an empirical $L_{\rm CO(1-0)}/L_{\rm H\alpha}$ radial profile, for which high ratios indicate enhanced CO luminosities and/or depressed H$\alpha$ luminosities (vice-versa for low ratios). Each data point shows the average $L_{\rm CO(1-0)}/L_{\rm H\alpha}$ ratio of an annulus of width $200$~pc centred on the galaxy centre, with a step size of $100$~pc, where the average $L_{\rm CO(1-0)}/L_{\rm H\alpha}$ is calculated as the ratio of the sum of $L_{\rm CO(1-0)}$ and the sum of $L_{\rm H\alpha}$ within the annulus. On the right-hand ordinate, we show the depletion time $\tau_{\rm dep}$ calculated by converting $L_{\rm CO(1-0)}$ to molecular gas mass and $L_{\rm H\alpha}$ to SFR. The up-pointing triangles indicate $\tau_{\rm dep}$ calculated using all the H$\alpha$ emission, irrespective of the ionisation mechanism. These are thus truly lower limits, as the SFRs are upper limits. For the three spiral galaxies in our sample, which contain reliably identified SF regions, we use down-pointing triangles for the $\tau_{\rm dep}$ calculated using only the H$\alpha$ emission reliably classified as star forming. This results in $\tau_{\rm dep}$ upper limits, as the SFRs are lower limits (all the regions where SF is potentially present but is not dominant are excluded). For those three spirals, the true $\tau_{\rm dep}$ are thus somewhere between the up- and the down-pointing triangles.

SF regions, where $\tau_{\rm dep}$ can be reliably measured, are present in only the three spiral galaxies. For these regions (indicated by down-pointing triangles in Figure~\ref{fig:dept_rad}),  the $\tau_{\rm dep}$ radial profiles all increase with decreasing radius. There is no region dominated by SF within a $\approx500$~pc radius in any galaxy. 

Focussing on the $\tau_{\rm dep}$ lower limits (using all of the H$\alpha$ emission as a SFR tracer), there is a large scatter of $\tau_{\rm dep}$ ($L_{\rm CO(1-0)}/L_{\rm H\alpha}$) among our sample galaxies. 
The majority of ETGs in our sample (NGC~524, NGC~708, NGC~4429 and NGC~4435) have flat or slightly increasing $L_{\rm CO(1-0)}/L_{\rm H\alpha}$ with decreasing radius. Both the H$\alpha$ and the CO flux are centrally concentrated (as illustrated in Figure~\ref{fig:maps}), while their ratios remain roughly constant as a function of galactocentric radius. This offers further evidence that the ionised gas and molecular gas phases are correlated. 
In NGC~383, NGC 4438 and NGC 4501, $L_{\rm CO(1-0)}/L_{\rm H\alpha}$ decreases with decreasing radius, especially within a galactocentric radius of $500$~pc. This indicates that there is a surplus of ionised-gas emission, associated with strong nuclear (AGN) activity, as summarised in Table~\ref{tab2}. 
NGC~3169 is the only galaxy with a sharp increase of $L_{\rm CO(1-0)}/L_{\rm H\alpha}$ as a function of decreasing radius. This galaxy has a rich molecular gas reservoir in the bulge with scarce SF and no AGN-dominated emission-line region. A possible mechanism preventing the molecular gas from forming stars is turbulence, induced by inflow from the SF ring into the bulge (see \citealt{2022MNRAS.514.5035Lu} for a detailed analysis and discussion).

\section{Discussion}
\label{discussion}

Summarising our results, we have established that SF is suppressed in our sample of ETGs and bulges. This is supported by the ionisation mechanism classification, the $\Sigma_{\rm SFR}$ -- $\Sigma_{\rm mol}$ scaling relations and the radial profiles of depletion time. 

Using the BPT diagram (see Section~\ref{BPT}), we showed that there is no region dominated by SF ionisation in our ETGs, despite them being cold molecular gas-rich, while the regions dominated by SF in our spiral galaxies are outside of their bulges. Using the WHAN diagram (see Section~\ref{BPT}), we further confirmed that our ETGs are retired galaxies, with very limited AGN ionisation; the gas is primarily ionised by old stars. The Seyfert ionisation and LINER ionisation of our sample galaxies are confined to their innermost regions. 

The lack of any SF region in our ETGs and bulges compromises our ability to trace SF using H$\alpha$ emission. The SFRs calculated using all of the H$\alpha$ emission are as a consequence upper limits, and hence set depletion time lower limits. Assessing these lower limits in terms of the resolved $\Sigma_{\rm SFR}$ -- $\Sigma_{\rm mol}$ relations (Section~\ref{KS}), spiral galaxies have large scatters, due to the mix of several ionisation mechanisms. Somewhat surprisingly, however, ETGs have tighter scaling relations. 
As these are $\tau_{\rm dep}$ lower limits, the implication is that SF is indeed quenched in those ETGs. If considering only reliably-classified SF regions, present only in our spiral galaxies, the $\Sigma_{\rm SFR}$ -- $\Sigma_{\rm mol}$ relation is consistent with that of nearby LTGs. The lower-limit $\tau_{\rm dep}$ radial profiles vary across our sample galaxies in both shape (centrally rising, centrally declining or approximately flat in the inner $600$~pc) and magnitude ($0.1$ to $300$~Gyr), but the majority of our ETGs and bulges have rising $\tau_{\rm dep}$ as a function of decreasing radius.

The questions therefore now are, how do we explain the depletion times of these ETGs and bulges? Can we relate those depletion times to some mechanisms that are regulating SF? What is common among these galaxies?

\subsection{SF regulation}
\label{dis:SF_reg}

Many SF regulation mechanisms have been proposed, including bulge dynamics, shear, stellar feedback and AGN feedback. Multiple mechanisms can of course be at play in any individual galaxy \citep[e.g.][]{2018NatAs...2..695Man}. Here, we briefly summarise the mechanisms that could be responsible for quenching SF in our sample galaxies.

\subsubsection{Bulge dynamics and shear}
\label{dis:bulge}

As shown by \citet{Davis2014MNRAS.444.3427D}, bulge dynamics is the favoured SF regulation mechanism in ETGs. In their sample of ETGs, the galaxies with the longest depletion times are those where most of the molecular gas is confined to the regions within which the rotation curves are still rising. The galaxies with the strongest suppression of SF also have the fastest rising rotation curves. Conversely, the galaxies with substantial gas in the flat part of the rotation curves have depletion times consistent with those of normal spiral galaxies \citep{2024Galax..12...36Ruffa}. 

\begin{figure*}
  \centering
  \includegraphics[width=0.9\textwidth]{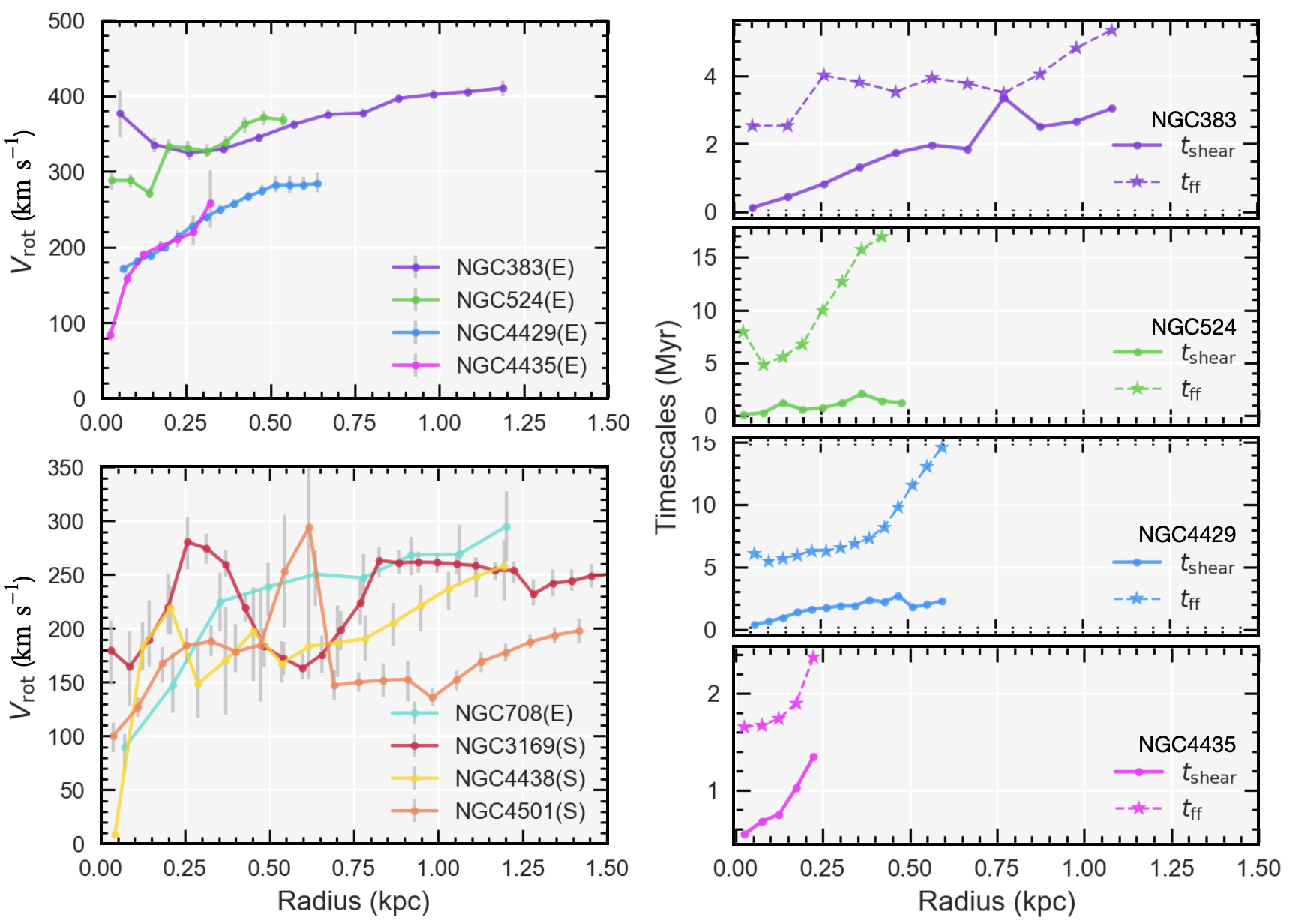}
  \caption{\label{fig:Vrot} Left: rotation velocities ($V_{\rm rot}$) of our sample galaxies as a function of galactocentric radius. The rotation velocities were derived from our ALMA CO data cubes using {\tt 3DBarolo}. Right: Comparison of the shear and free-fall timescales of our four galaxies with regularly-rotating molecular gas discs. 
  }
\end{figure*}

In Figure~\ref{fig:Vrot}, we show the molecular gas rotation velocity ($V_{\rm rot}$) as a function of the galactocentric radius for each of our sample galaxies. To generate these rotation curves, we use the {\tt 3DBarolo} software\footnote{https://editeodoro.github.io/Bbarolo/} of \citet{teodoro20153d} to fit tilted-ring models to our three-dimensional (3D) CO emission-line data cubes, taking advantage of the high spectral and spatial resolutions of our ALMA data. We use the ALMA data cubes at their native angular resolutions, as described in \citet{Davis2022MNRAS.512.1522D}, and the radii of the tilted rings are set to $1.5$ times the synthesised beam sizes. For each galaxy, following a first fit with all parameters free, the dynamical centre (spatially and spectrally), position angle and inclination are then fixed.

In the top-left panel of Figure~\ref{fig:Vrot}, we show the rotation curves of NGC~383, NGC~524, NGC~4429 and NGC~4435. These four galaxies have regularly rotating CO discs with neglegible non-circular motions. Thus, the rotation curves can be reliably obtained using {\tt 3DBarolo}. NGC~383 and NGC~524 have very high $V_{\rm rot}$ at their centres, gently rising with radius to reach $\approx400$~km~s$^{-1}$. NGC~4429 and NGC~4435 have low central $V_{\rm rot}$ of $\lesssim100$~km~s$^{-1}$ but rapidly-rising rotation curves that reach $\gtrsim300$~km~s$^{-1}$. Overall, the rotation velocities of these four ETGs reach higher maximum rotation velocities than those of nearby LTGs. Many of these LTGs have maximum rotation velocities $V_{\rm rot,max}\approx150$~km~s$^{-1}$, while the earlier LTGs (with earlier Hubble types and higher stellar masses) have $V_{\rm rot,max}\approx200$~km~s$^{-1}$ \citep{2020ApJ...897..122Lang}. The high rotation velocities at the centre of NGC~383 are due to the prominent Keplerian motions surrounding its central SMBH \citep{2025MNRAS.tmp...59Z, 2019MNRAS.490..319North}.

The strong shear associated with the large velocity gradients observed is potentially important for the regulation of the molecular gas properties and SF. This can be quantified by calculating and comparing the shear ($\tau_{\rm shear}$) and free-fall ($\tau_{\rm ff}$) timescales using the following equations:
\begin{equation}
      \tau_{\rm shear}\equiv\tau_{\rm shear}(R)  =\frac{1}{2A(R)}, 
\end{equation}
\begin{equation}
   \tau_{\rm ff}\equiv\tau_{\rm ff}(R)=\sqrt{\frac{3\pi H}{32G\Sigma_{\rm gas,disc}(R)}}\,\,\,,
\end{equation}
where $A(R)\equiv-\frac{R}{2} \frac{d\Omega(R)}{dR}|_R$ is Oort's constant $A$ evaluated at the galactocentric radius $R$, $\Omega(R)\equiv V_{\rm rot}(R)/R$ is measured, $H$ is the line-of-sight depth of the molecular gas layer, taken here to be $100$~pc based on the characteristic thickness of the molecular gas layer of the MW and other galaxies \citep{2013ApJ...779...43Pety, 2014AJ....148..127Yim, 2015ARA&A..53..583Heyer}, and $\Sigma_{\rm gas,disc}$ is the azimuthally-averaged molecular gas mass surface density of the disc within each tilted ring of the {\tt 3DBarolo} model. We show the comparison of $\tau_{\rm shear}$ and $\tau_{\rm ff}$ in the right panel of Figure~\ref{fig:Vrot} for the four galaxies with regularly rotating CO discs. The $\tau_{\rm shear}$  are significantly shorter than $\tau_{\rm ff}$ in all cases. This indicates that shear is the dominant mechanism regulating the molecular gas behaviour of these four galaxies, as the mechanism with the shorter timescale must dictate the structures and dynamics of the molecular gas clouds.

In the bottom-left panel of Figure~\ref{fig:Vrot}, we show the rotation curves of NGC~708, NGC~3169, NGC~4438 and NGC~4501. These four galaxies have more complicated kinematics, and thus the derived $V_{\rm rot}$ have larger uncertainties. The three spiral galaxies in our sample (NGC~3169, NGC~4438 and NGC~4501)  all have rising rotation curves with increasing radius, that then drop and flatten at the transitions between the bulges and the discs. There may exist multiple sources of non-circular motions, including outflows from the AGN and/or inflows from the spiral arms into the bulges and warped discs. NGC~708 has consistently rising $V_{\rm rot}$ as a function of radius. However, these $V_{\rm rot}$ are not constrained very well, because the molecular (and ionised) gas of NGC~708 has clear filamentary structures (likely the result of hot halo cooling and gas infall, that are common in BCGs; \citealt{2021MNRAS.503.5179North,2024Galax..12...36Ruffa}).

Evidence of the bulge dynamics regulating SF is also present in the morphologies of the ISM. As shown in Figure~\ref{fig:maps}, our sample ETGs and bulges have smooth molecular gas discs and diffuse ionised gas. 
\citet{Davis2022MNRAS.512.1522D} quantified the ISM morphology using non-parametric morphological indicators (Gini, Smoothness and Asymmetry). They demonstrated that ETGs have smoother and more rotationally symmetric central gas discs than LTGs. Furthermore, the non-parametric morphological indicators are correlated with the central stellar mass surface density ($\mu_\star$). Galaxies with larger $\mu_\star$ also have smoother discs.
The $\mu_\star$ of our galaxies are listed in Table~\ref{tab1} and are all larger than those of typical star-forming LTGs (for which $\log(\mu_\star/{\rm M_\odot\,kpc^{-2}})\approx8.5$, \citealt{Davis2022MNRAS.512.1522D}). This correlation implies that the deep potential wells of bulges affect the ISM morphology and thus the SF activity of ETGs. Simulations of gas discs in bulges have also shown that bulge dynamics can regulate the ISM morphology and SFR \citep{gensior2023wisdom}. 

A few of our targets have also been subject to more detailed analyses of their bulge dynamics. \citet{2022MNRAS.514.5035Lu} showed that the rising radial profile of depletion time (with decreasing radius) of NGC~3169 is consistent with that generated from numerical simulations including a large bulge \citep{gensior2020heart, gensior2021elephant}. Thus, the existence of a bulge can partially explain the quenched SF at the centre of NGC~3169. In NGC~4429, \citet{Liu2021MNRAS.505.4048L} showed that strong shear, created by the deep and steep gravitational potential of the galaxy, is responsible for the high virial parameters ($\alpha_{\rm vir}$) of the molecular clouds. A high $\alpha_{\rm vir}$ implies that the internal gravity of a cloud is overpowered by the internal (random) motions and/or forces external to the cloud. In NGC~4429, shear from the large-scale rotation of the disc is significant and prevents the clouds from becoming self-gravitating (the clouds identified are thus most likely in fact transient gas overdensities). 
In NGC~524, shear is once again responsible for high molecular cloud virial parameters \citep{2024MNRAS.531.3888Lu}. In fact, the only molecular clouds that could survive shear in NGC~524 are much smaller than the structures detected. 

All of the above stresses the importance of bulge dynamics and the strong shear associated with spheroids for the regulation of SF. 

\subsubsection{AGN feedback}
\label{dis:AGN}

All of our sample galaxies host an AGN or some type of nuclear activity, summarised in Table~\ref{tab2}. 
However, the effectiveness of AGN feedback to quench SF is inconclusive. As discussed in Section~\ref{BPT}, the gas ionisation of our ETGs generally results from old stellar populations, with Seyfert and LINER ionisation confined at most to the central few hundred parsecs in radius. 

AGN feedback can sometime suppress SF by propelling material out of the galaxies and/or keeping the gas hot \citep[see e.g.\ the review by][]{morganti2017many}. Our ETGs and bulges are rich in molecular gas, indicating that their AGN are not efficient at ejecting material. AGN can also drive outflows that enhance turbulence in molecular gas discs and suppress SF. In our sample galaxies, the molecular and ionised-gas discs are mostly smooth, but non-circular motions (e.g., warped discs) and AGN-driven outflows are potentially present in some of our sample galaxies, especially NGC~383 (Zhang et al., in prep.), NGC~708 \citep{2021MNRAS.503.5179North} and NGC~3169 \citep{2022MNRAS.514.5035Lu}. However, their role in regulating SF beyond the nuclear regions is unclear. 
Investigations of AGN feedback in nearby galaxies have indicated that, in the presence of a jet-driven AGN outflow, there may be strong [\ion{O}{iii}] emission and high H$\alpha$ velocity dispersions 
perpendicular to the direction of the jets  \citep{venturi2021magnum, gao2021nuclear}, but there is no clear evidence of such features in our sample galaxies. 

One particularly interesting case is NGC~3169, where there is intense SF in the ring just outside the bulge. The regions surrounding the SF peaks are classified as composite by the BPT diagram and Seyfert-ionised by the WHAN diagram, indicating AGN radiation within a background of SF regions. This kind of radiation can sometimes result in positive AGN feedback, enhancing SF \citep[see an analogous study in e.g.][]{pak2023origin}. 

There is no evidence of direct suppression of SF by AGN feedback in our sample galaxies, although this does not necessarily mean that AGN feedback is not important across the ETG and bulge population. A larger sample of ETGs harbouring an AGN is required to clearly disentangle which SF-regulation mechanism is dominant, as this will likely differ for each galaxy (depending on the stellar potential and AGN properties).

\subsubsection{Stellar feedback}
\label{dis:turb}

Stellar feedback, and the induced turbulence, has also been proposed and thoroughly investigated as a candidate to suppress SF \citep[e.g.][]{2007ARA&A..45..565McKee}. However, the effectiveness of stellar feedback is questionable in ETGs and bulges, as these systems have deep gravitational potential wells and are dominated by old stars. In ETGs and bulges, the timescale for stellar feedback ($\approx4$~Myr post SF for type II supernovae) is also inconsistent with the time difference between the past active SF and the present quenched SF (of the order of Gyrs). As clearly shown by the maps of Figure~\ref{fig:maps}, the molecular gas discs are smooth and the ionised gas is diffuse. The lack of molecular gas clumps and/or ionised-gas bubbles indicates that stellar feedback is currently at best limited. 

The spiral galaxies in our sample are more complicated. Their bulges are similar to the ETGs, with smooth molecular gas discs and no SF-dominated region. Immediately outside the bulges of NGC~4501 and NGC~3169, however, star-forming rings (or the innermost parts of spiral arms) are present. There are many SF regions with high SFRs and potentially strong stellar feedback in these structures, potentially explaining the large scatter of the $\Sigma_{\rm SFR}$ -- $\Sigma_{\rm mol}$ relations. 

\subsubsection{Galaxy interactions}
\label{dis:inter}

Our sample contains a peculiar interacting pair of galaxies located in the Virgo cluster of galaxies, NGC~4438 and NGC~4435, also called "The Eyes". NGC~4435 has a typical ETG morphology. Its molecular gas and ionised-gas discs, ionisation mechanisms and depletion times are all similar to those of the other ETGs in our sample. However, the ionised-gas emission lines are faint compared to the continuum. NGC~4438 has a severely-distorted ionised-gas disc and nuclear activity likely associated with an AGN \citep{2007MNRAS.380.1009Hota4438, 2022MNRAS.515.2483Li4438}. This mixture of different ionisation mechanisms results in a large $\tau_{\rm dep}$ scatter. Nevertheless, separating the ionisation sources, there is only one region dominated by SF in this interacting pair, $\approx200$~pc away from the centre of NGC~4438. The $\tau_{\rm dep}$ of this region is consistent with those of the SF regions in the other two spiral galaxies in our sample. 

Analogously to AGN feedback, galaxy interactions and mergers can act to both enhance and suppress SF. Interacting pairs are thus a good laboratory to study the physical conditions that can drive or prevent SF.

\subsection{Sample bias and uncertainties}
\label{dis:bias}

\subsubsection{Sample size}
\label{dis:sample}

Our sample contains all the galaxies that have relatively high spatial resolution ALMA and SITELLE observations. Potential biases can therefore arise from the limited size of the sample, $8$ galaxies. Each of these eight galaxies has its own unique features and a specific type of AGN, summarised in Tables~\ref{tab1} and \ref{tab2}. We focussed on correlating the shared properties of the galaxies (e.g.\ bulges) with the similarities of their SFEs. However, we note that a larger sample is clearly necessary to better disentangle some of the SF-regulation mechanisms and draw statistically-meaningful conclusions. 

\subsubsection{SFR tracer}
\label{dis:SFR}

In this work we have used H$\alpha$ emission as a measure of SFR. Using only this tracer implies that we are only focussing on SF due to massive stars formed $\lesssim10$~Myr ago \citep{kennicutt2012star}. However, the fraction of recently-formed massive stars in ETGs might be different from that in LTGs. Recent hypotheses have suggested that galaxies with old stellar populations and high metallicities can form stars efficiently, but are not able to form the O and B stars that generate ionised-gas bubbles and hence H$\alpha$ emission \citep{2024arXiv240203423SnoOB}.

Estimating SFRs is also particularly difficult when using high-spatial resolution observations. As discussed in Section~\ref{data:Halpha}, the conversion from H$\alpha$ flux to SFR may break down at spatial scales smaller than $\approx500$~pc, due to the variations of the SF histories and metallicities below this scale \citep{kennicutt2012star}. H$\alpha$ fluxes are also affected by dust extinction. Although we corrected for extinction using the H$\beta$ fluxes, there are a large number of spaxels for which H$\beta$ is detected with only very low $S/N$, and thus for which the extinction was estimated based on that of other nearby spaxels. 

Another significant argument for SF quenching presented in this work comes from the ionisation-mechanism classification: we demonstrated that there is no region where SF dominates the ionisation in any of our sample ETGs and bulges. 
Alternative tracers would be required to measure SF from lower-mass stars in these systems. One option is to use FUV+$22~\mu$m measurements, as $22~\mu$m in particular is more sensitive to lower mass stars. 

\subsubsection{CO-to-H$_2$ conversion}
\label{dis:Xco}

In this work we adopted the MW CO-to-molecules conversion factor $X_{\rm CO(1-0)}$, that is also routinely applied to other nearby galaxies \citep[e.g.][]{sun2018cloud, Liu2021MNRAS.505.4048L}. A fixed $X_{\rm CO(1-0)}$ was adopted for all the galaxies in our sample to ensure fair comparisons. However, $X_{\rm CO(1-0)}$ can vary as a function of metallicity and in different environments (see e.g.\ \citealt{bolatto2013co} for a review). Recent spatially-resolved investigations of nearby galaxies have revealed that $X_{\rm CO(1-0)}$ can be smaller in galaxy centres than in galaxy discs \citep{2013ApJ...777....5Sandrom, 2022ApJ...925...72Teng,2022MNRAS.516.4066Lelli}. Such a radial dependence of $X_{\rm CO(1-0)}$ could therefore (at least partially) explain the observed trend of increasing $\tau_{\rm dep}$ with decreasing galactocentric radius. However, due to a lack of $X_{\rm CO(1-0)}$ measurements in ETGs, whether $X_{\rm CO(1-0)}$ will also decrease with radius in ETGs with steep and deep (and nearly spherical) potential wells is unknown. Further work on constraining $X_{\rm CO(1-0)}$ in such environments is necessary to constrain this possibility. 
We note that the uncertainties due to potential variations of $X_{\rm CO(1-0)}$ are not included in the uncertainties reported in this work, although in general an uncertainty of $\approx0.3$~dex should be added to all the molecular gas mass measurements when a constant $X_{\rm CO(1-0)}$ is adopted \citep{bolatto2013co}.

\section{Conclusions}
\label{conclusion}

We have used ALMA and SITELLE observations to study the molecular and ionised gas of a sample of 8 ETGs and bulges of spiral galaxies, quantify their spatially-resolved depletion times and identify the mechanisms that regulate their SF. Our main findings are summarised below.

\begin{enumerate}
\item We do not identify any SF-dominated region in the BPT and WHAN diagram of any of our ETGs and bulges (i.e.\ within the inner $500$~pc in radius of spiral galaxies). In the spiral galaxies, there are some regions dominated by SF ionisation just outside the bulges, in the innermost regions of the spiral arms. 

\item Our ETGs and bulges have ionised gas arising mostly from old stars, despite some AGN ionisation at the very centres. 

\item The $\Sigma_{\rm SFR}$ -- $\Sigma_{\rm mol}$ relations of ETGs (derived using SFR upper limits) are tight. The slopes are consistent with the relations of nearby spiral galaxies, while the scatter is smaller. The $\Sigma_{\rm SFR}$ --$ \Sigma_{\rm mol}$ relations of our spiral galaxies have large scatters, caused by the mixture of ionisation mechanisms. 

\item The radial profiles of depletion time ($\tau_{\rm dep}$) of our ETGs (again derived using SFR upper limits) reveal increasing $\tau_{\rm dep}$ with decreasing radius. The SF regions of our spiral galaxies have a similar trend. This suggests inside-out quenching in our sample galaxies.

\item We show that bulge dynamics (particularly shear due to deep and steep gravitational potential wells) is an important SF-regulation mechanism in at least half of our sample galaxies. We also explored other factors including AGN feedback, stellar feedback and galaxy interactions.

\item New methods to estimate SFRs and better constrain the CO-to-molecules conversion factors $X_{\rm CO}$ of ETGs are crucial to more accurately probe the SFEs of ETGs and bulges.
\end{enumerate}

\section*{Acknowledgements}

This research is based on observations obtained with the SITELLE instrument on the Canada-France-Hawaii Telescope (CFHT) which is operated from the summit of Maunakea, and the Atacama Large Millimeter/submillimeter Array (ALMA) in the Atacama desert.
This paper makes use of the following ALMA data: ADS/JAO.ALMA\#013.1.00493.S, ADS/JAO.ALMA\#2015.1.00419.S, ADS/JAO.ALMA\#2015.1.00466.S, ADS/JAO.ALMA\#2015.1.00598.S, ADS/JAO.ALMA\#2016.1.00437.S, ADS/JAO.ALMA\#2016.2.00053.S, ADS/JAO.ALMA\#2017.1.00391.S and ADS/JAO.ALMA\#2019.1.00582.S. ALMA is a partnership of ESO (representing its member states), NSF (USA) and NINS (Japan), together with NRC (Canada), MOST and ASIAA (Taiwan), and KASI (Republic of Korea), in cooperation with the Republic of Chile. The Joint ALMA Observatory is operated by ESO, AUI/NRAO and NAOJ. The National Radio Astronomy Observatory is a facility of the National Science Foundation operated under cooperative agreement by Associated Universities, Inc. The authors acknowledge support from the Centre de recherche en astrophysique du Québec, un regroupement stratégique du FRQNT.

We are grateful to the CFHT and ALMA scheduling, data processing and archive teams. We also wish to acknowledge that the summit of Maunakea is a significant cultural and historic site for the indigenous Hawaiian community, while the the high-altitude plateau Chajnantor on which the ALMA telescope sits is sacred to indigenous Likanantai people. We are most grateful to have the opportunity of observing there. We thank Federico Lelli for help with the {\tt 3DBarolo} analyses, Alberto Bolatto for valuable comments and Nicole Ford for scientific and aesthetic support. 

AL, HB, DH, CR, LD and LRN acknowledge funding from the NSERC Discovery Grant and the Canada Research Chairs (CRC) programmes. MB was supported by STFC consolidated grant `Astrophysics at Oxford' ST/H002456/1 and ST/K00106X/1. JG gratefully acknowledges funding via STFC grant ST/Y001133/1 and financial support from the Swiss National Science Foundation (grant no CRSII5\_193826). 
FHL acknowledges support from the ESO Studentship Programme.
TAD acknowledges support from the UK Science and Technology Facilities Council through grant ST/W000830/1. 

\section*{Data Availability}
The raw data underlying this article are publicly available on the National Radio Astronomy Observatory (programmes 2013.1.00493.S, 2015.1.00466.S, 2015.1.00419.S, 2015.1.00598.S, 2016.1.00437,
2016.2.00053.S, 2017.1.00391.S and
2019.1.00582.S) and CFHT archives (programmes 20BC09, 20BC25, 22BC99, 23AC06 and 24AC18). All analysed data products can be found in the WISDOM data archive\footnote{https://wisdom-project.org/data/}.

\bibliographystyle{mnras}
\bibliography{ref_v3}

\appendix

\section{SITELLE data analysis: stellar continuum subtraction}
\label{app_continuum}
 
By default, when fitting the emission lines, the SITELLE data analysis pipelines ({\tt ORCs} and {\tt Luci}) model the continuum at each spaxel as a constant ("Flat" model). In our sample galaxies, it is however necessary to consider the background stellar population spectrum at each spaxel, as the absorption lines of the stellar continuum can affect the derived emission-line fluxes. Due to the limited sensitivity of the SITELLE spectra, we need to bin the spectra over a large region to achieve the $S/N$ necessary for a reasonable stellar population spectrum fit. For each galaxy, we therefore test two additional methods: 1) we bin $100\times100$ spaxels centred on the galaxy centre, referred to as the ``Integrated'' model; and 2) we bin spaxels within annuli of width $150$~pc centred on the galaxy centre (with mean radii ranging from $0$ to $900$~pc and a step size of $150$~pc), referred to as the ``Ring'' model. 
The motivations for adopting these two additional methods are explained in detail in \citet{2024MNRAS.531.3888Lu}. For each binning method, we de-redshift the spectrum at each spaxel using a luminosity-weighted mean line-of-sight velocity map obtained from the initial fit of the emission lines (adopting a flat continuum). 

We use the {\tt pPXF} algorithm \citep{2023MNRAS.526.3273Cappellari} with the Medium-resolution Isaac Newton Telescope Library of Empirical Spectra (MILES; \citealt{sanchez2006MNRAS.371..703S}) to fit the binned spectra.
We combine the SN3 and SN2 data cubes to maximise the wavelength coverage and perform an interpolation to adjust the spectral sampling to a constant step, without changing the spectral resolution of each filter. 
As the two most prominent absorption components within the wavelength range of our observations are the H$\alpha$ and H$\beta$ absorption lines, we do not mask the emission lines of each spectrum. Instead, we mask spaxels with strong emission by setting a threshold on the [\ion{N}{ii}] emission lines (typically $S/N>3$). 
To account for SITELLE’s sinc line spread function, we modify {\tt pPXF} by replacing its emission line Gaussian profile by a sinc profile. As the broadening of the absorption features in the SITELLE spectra is nevertheless well represented by a Gaussian, we use the MILES library spectra as they are. We fit the emission and absorption lines simultaneously and adopt a multiplicative Legendre polynomial with a degree of $4$. We do not interpret the best-fitting MILES model spectra physically, but merely use them to model and subtract the continua and remeasure the emission line fluxes using the continuum-subtracted spectra.

\begin{figure*}
\centering\includegraphics[width=0.95\textwidth]{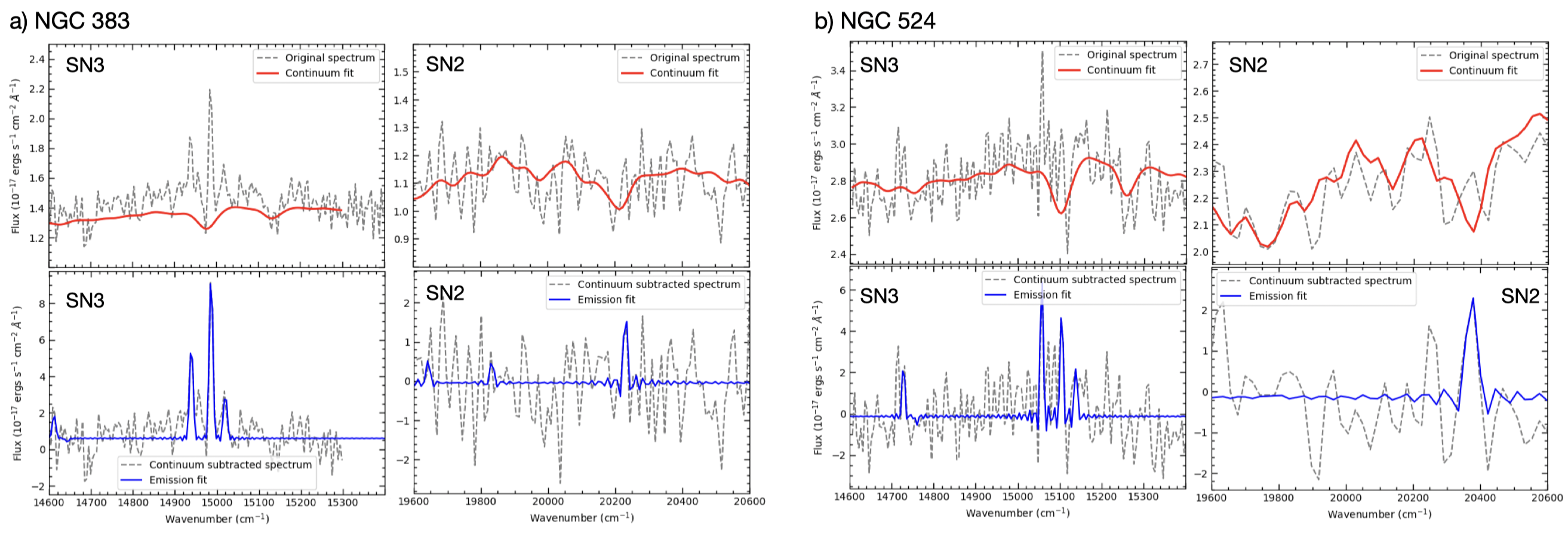}\\\includegraphics[width=0.95\textwidth]{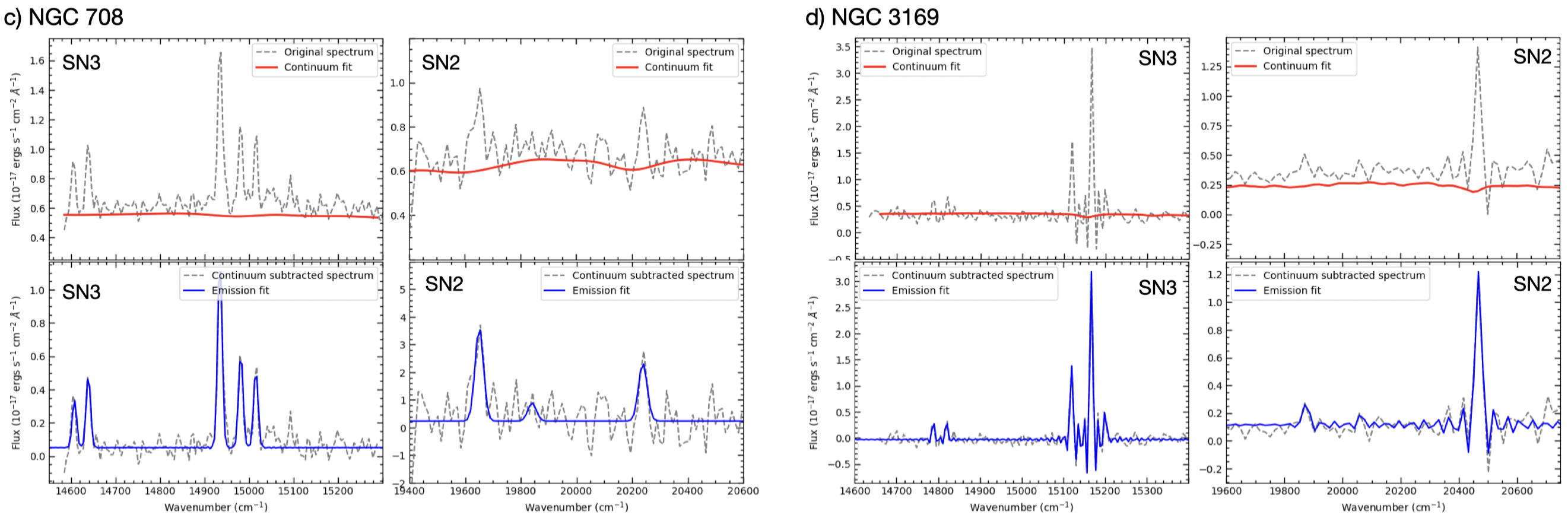}\\\includegraphics[width=0.95\textwidth]{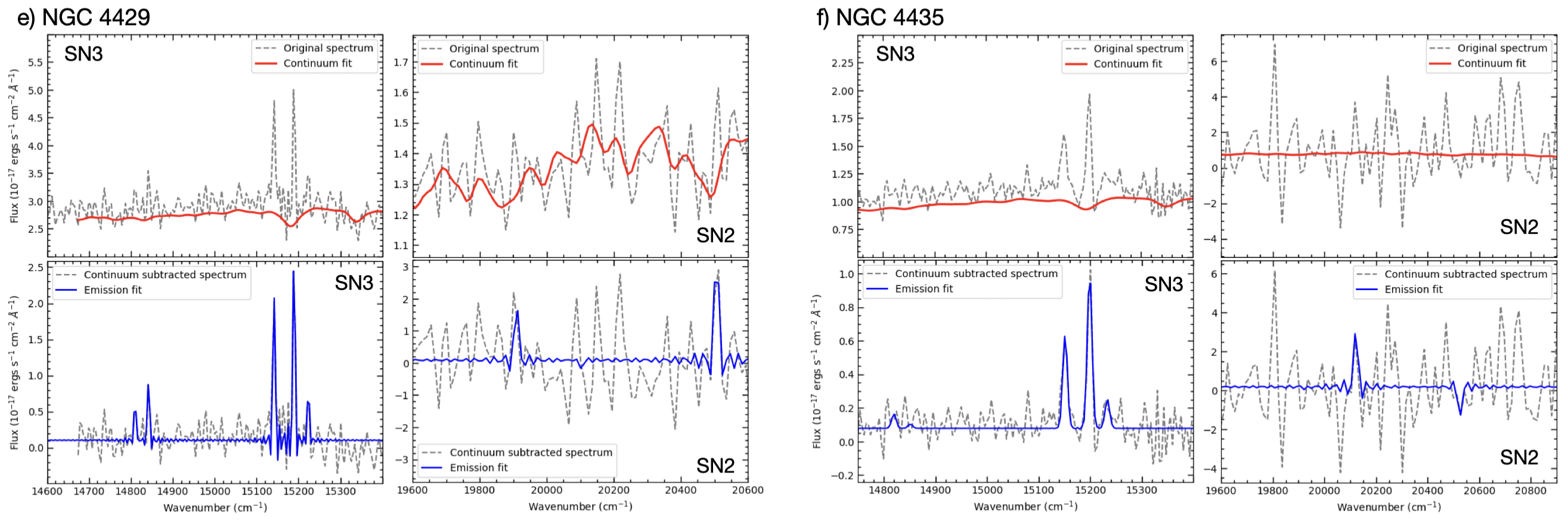}\\\includegraphics[width=0.95\textwidth]{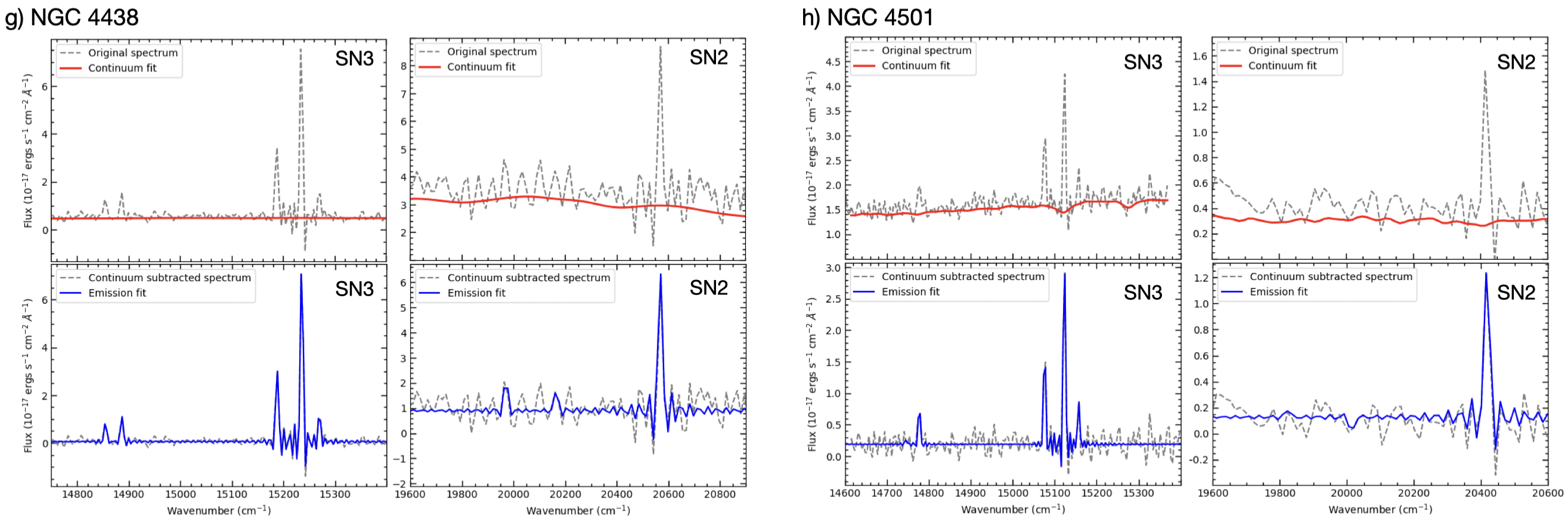}
  \caption{Examples of SITELLE spectra from both the SN3 and the SN2 data cubes. In each panel, we show spectra extracted from a spaxel representative of the typical $S/N$ across the data cubes. The top row shows the original spectra (grey dashed lines) and the best-fitting stellar continua (red solid lines) using {\tt ORCS}. The bottom row shows the continuum-subtracted spectra (grey dashed lines) and the best-fitting emission-line model (blue solid lines).
  }
  \label{fig:app1}
\end{figure*}

For each galaxy, we test the ``Integrated'' and the ``Ring'' models by fitting the emission lines and measuring the spectrum RMS (i.e.\ the "noise" $\sigma_{\rm noise}$, measured with the emission lines masked) both before and after subtracting the best-fitting stellar population spectrum, which is rescaled to the spectrum at each spaxel. Specifically, we compare the average H$\alpha$ integrated flux difference ($\Delta F_{\rm H\alpha}$) and the average $\sigma_{\rm noise}$ before and after the stellar population spectrum subtraction. We list these statistics in Table~\ref{tabA1} and show an example of this process in Figure~\ref{fig:app1} for each galaxy. In all cases, $\sigma_{\rm noise}$ is smaller after subtracting the best-fitting stellar population spectrum, indicating that our {\tt pPXF} fits are modelling the stellar continua well. For NGC~524 and NGC~4429, the average $\Delta F_{\rm H\alpha}$ is significantly larger than the average $\sigma_{\rm noise}$ when using the ``Ring'' model, so we adopt this model to produce the emission-line moment maps. For NGC~708, NGC~3169, NGC~4438 and NGC~4501, which include all the spiral galaxies and the BCG of our sample, the average $\Delta F_{\rm H\alpha}$ is smaller than the average $\sigma_{\rm noise}$, so we consider the impacts of the absorption components to be negligible. For these galaxies, we thus adopt the best-fitting flat continuum ("Flat" model) to produce the emission-line moment maps. For NGC~383 and NGC~4435, the results of the ``Integrated'' model and the ``Ring'' model are approximately the same but the ``Integrated'' model is more efficient computationally, so we adopt the ``Integrated'' model. 

As shown in Figure~\ref{fig:app1}, small fractions of the continua often remain after subtracting the best-fitting stellar population spectra. When using {\tt ORCs} or {\tt Luci} to fit the emission lines, a flat continuum is thus also included in all cases to remove any possible remaining continuum emission.

\begin{table*}
\caption{Best-fitting stellar population spectrum subtraction statistics.}
\label{tabA1}
\centering
\begin{tabular}{|l|c|c|c|l|} \hline  
Target & $\Delta F_{\rm H\alpha}$& $\sigma_{\rm noise}$ before& $\sigma_{\rm noise}$ after& Model \\  
 & ($10^{-18}$~erg~s$^{-1}$~cm$^{-2}$)& ($10^{-18}$~erg~s$^{-1}$~cm$^{-2}$)& ($10^{-18}$~erg~s$^{-1}$~cm$^{-2}$)& \\  
 (1)& (2)& (3)& (4)& (5)\\ \hline 
NGC~\phantom{0}383 & 6.21& 1.61& 0.75& Integrated\\   
NGC~\phantom{0}524 & 9.08& 1.45& 1.25& Ring\\   
NGC~\phantom{0}708 & 2.50& 2.97& 2.56& Flat\\ 
NGC~3169 & 2.50& 6.85& 6.84& Flat\\   
NGC~4429 & 8.00& 1.89& 1.80& Ring\\   
NGC~4435$^\dagger$ & 15.0& 9.53& 8.31& Integrated\\   
NGC~4438$^\dagger$ & 1.00& 1.08& 1.07& Flat\\
NGC~4501 & 5.70& 1.78& 1.42& Flat\\ 
\hline 
\end{tabular}

{\it Notes:} {Summary of SITELLE data statistics before and after subtracting the best-fitting background stellar population spectrum. 
For NGC~524 and NGC~4429, the "Ring" model is used to calculate the statistics. For the other galaxies, the "Integrated" model is used. After evaluating the statistics, we then choose to proceed with the model listed in column (5). 
(1) Galaxy name. The interacting galaxy pair, NGC~4438 and NGC~4435, is marked with daggers. (2) Average difference between the integrated H$\alpha$ flux before and after subtracting the background stellar population spectrum. (3) Average RMS of the spectrum ($\sigma_{\rm noise}$) integrated over the average FWHM of the H$\alpha$ emission line before subtracting the background stellar population spectrum. (4) Average RMS of the spectrum ($\sigma_{\rm noise}$) integrated over the average FWHM of the H$\alpha$ emission line after subtracting the background stellar population spectrum. (5) Adopted stellar continuum model (see Appendix~\ref{app_continuum}).
}
\end{table*}

\section{Ionised-gas emission-line ratios}
\label{app_lineratios}

\begin{figure*}
  \centering\includegraphics[width=0.99\textwidth]{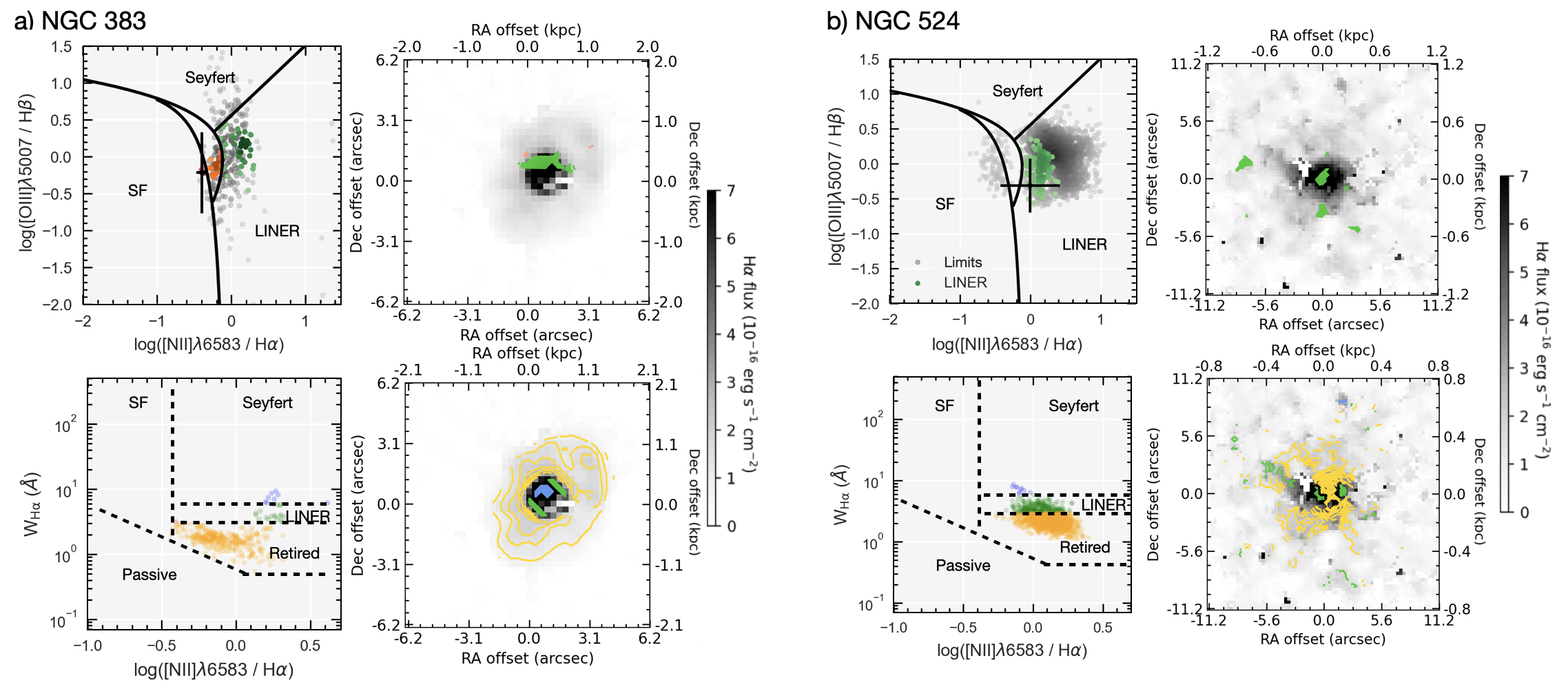}\\\includegraphics[width=0.99\textwidth]{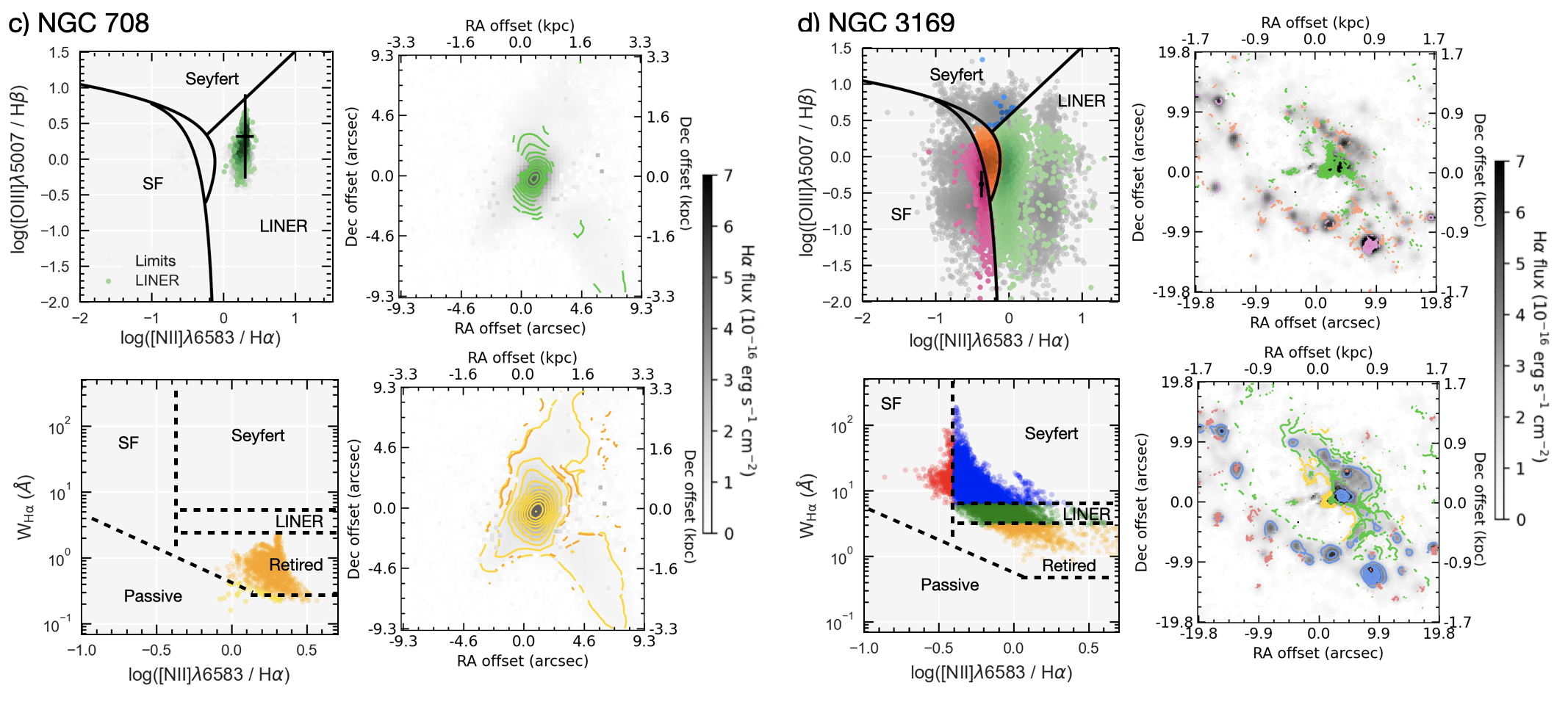}
  \caption{BPT diagram (top-left panel) and WHAN diagram (bottom-left panel), used to classify the ionisation mechanisms of NGC~383, NGC~524, NGC~708 and NGC~3169. The data are colour-coded according to the dominant ionisation source. The grey data points in the top-left panel indicate that one of the emission lines does not have $S/N>3$. The ionisation classification boundaries for the BPT diagram (solid black lines) and the WHAN diagram (dashed black lines) are taken from \citet{kewley2006host} and \citet{2011MNRAS.413.1687Cid}, respectively. The locations of the different ionisation sources are overlaid on the H$\alpha$ integrated flux maps using matching colours (top-right and bottom-right panels).
  }
  \label{fig:app2}
\end{figure*}

\begin{figure*}
  \centering\includegraphics[width=0.99\textwidth]{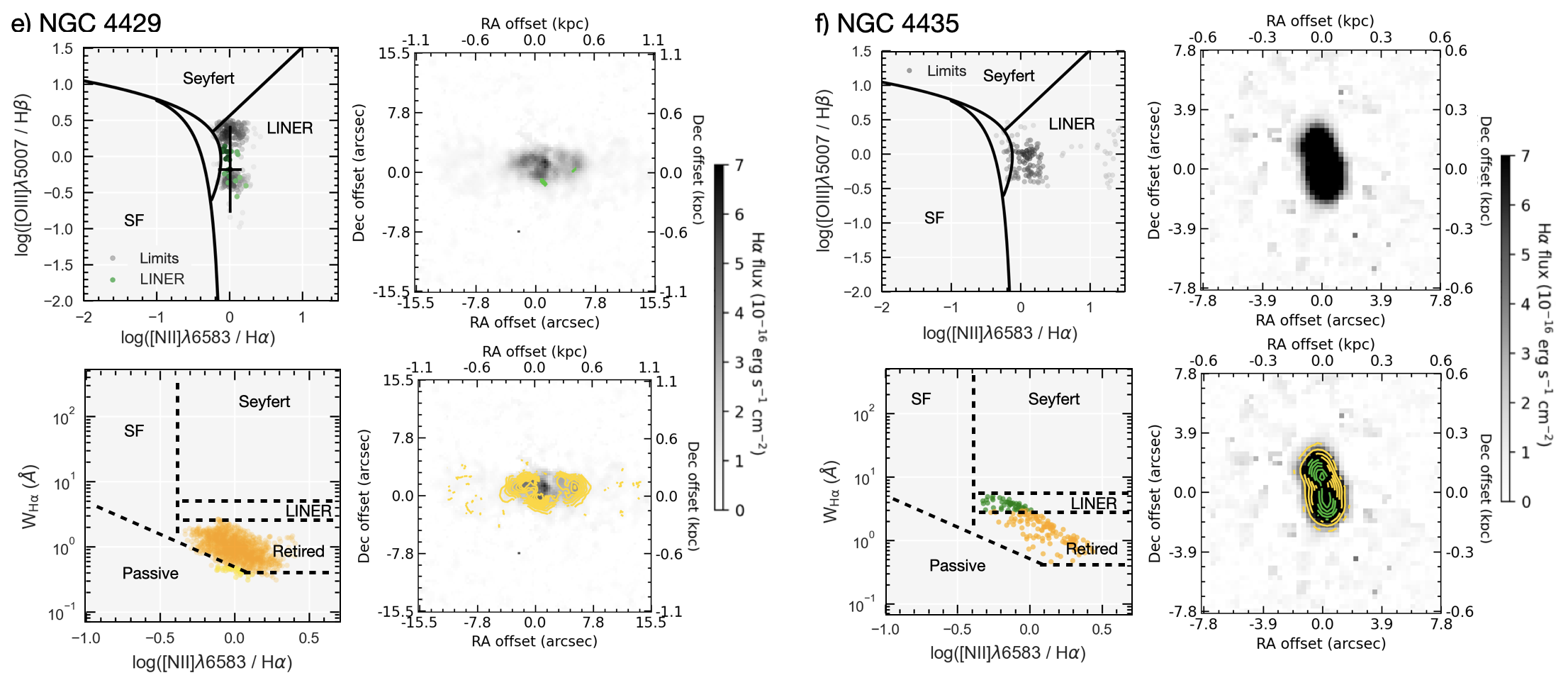}\\\includegraphics[width=0.99\textwidth]{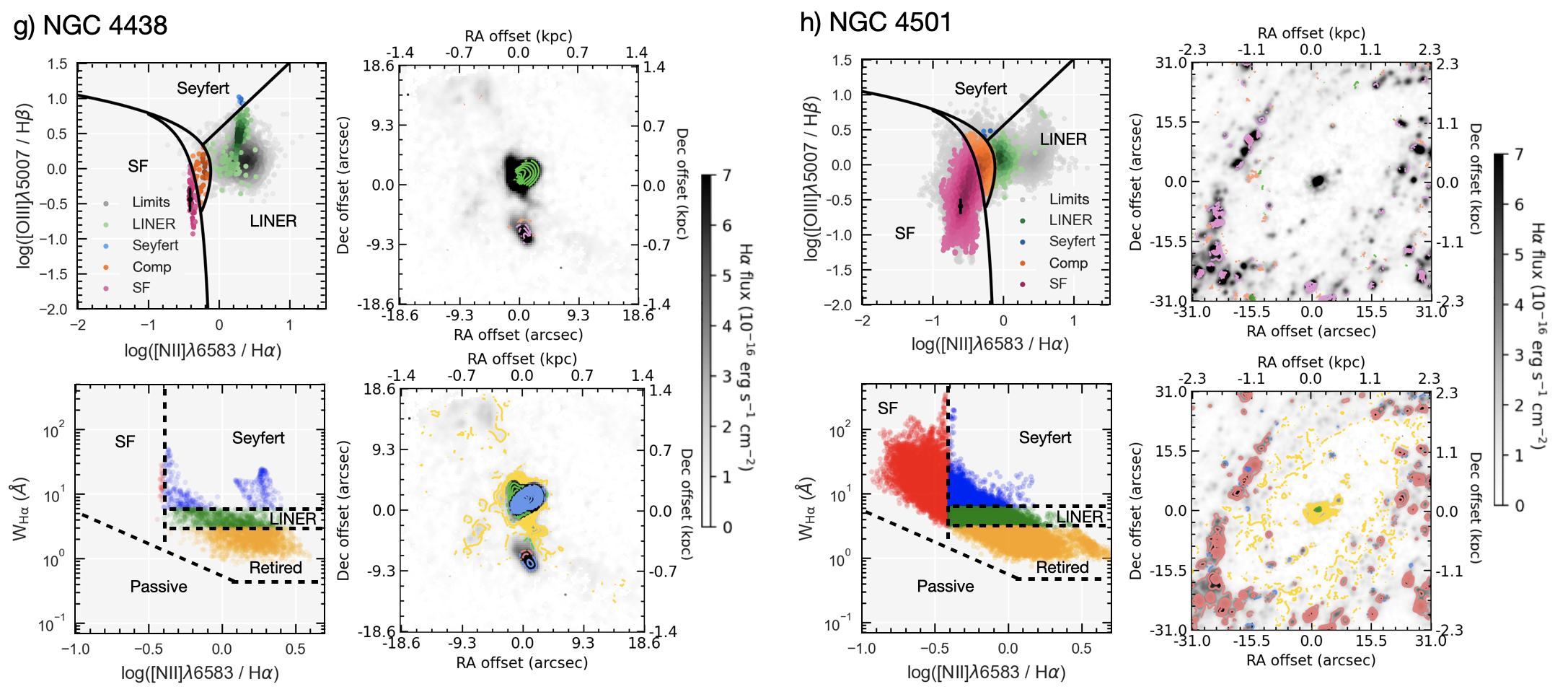}
  \caption{Same as Figure~\ref{fig:app2} but for NGC~4429, NGC~4435, NGC~4438 and NGC~4501}
  \label{fig:app22}
\end{figure*}

\end{document}